\documentclass[journal,draftcls,onecolumn,12pt,twoside]{IEEEtran}
\IEEEoverridecommandlockouts
\usepackage{tabularx}
\usepackage{cite}
\usepackage{verbatim}
\usepackage{amsmath,amssymb,amsfonts}
\usepackage{cuted}
\usepackage[noend]{algorithmic} 
\usepackage{textcomp}
\usepackage{xcolor}
\usepackage[caption=false,font=footnotesize]{subfig}
\usepackage{amsthm}
\usepackage{ulem}
\usepackage{enumitem}
\usepackage{graphicx} 
\usepackage{color, soul}
\usepackage{setspace}
\usepackage{cleveref}
\usepackage{multirow}
\usepackage{varwidth}
\usepackage{dblfloatfix}
\usepackage{balance}
\usepackage{chngcntr}
\usepackage{epstopdf}
\usepackage{layouts}
\usepackage{balance} 
\usepackage{amssymb}

\newcommand{\BS}{\textrm{BS}}
\newcommand{\chgain}{\tilde{h}}
\newcommand{\pfix}{P^{\text{fix}}_b}

\usepackage[breakable, skins]{tcolorbox}
\tcbset{arc=0mm,size=fbox,attach title to upper={\ ---\ },coltitle=black}

\def\BibTeX{{\rm B\kern-.05em{\sc i\kern-.025em b}\kern-.08em
    T\kern-.1667em\lower.7ex\hbox{E}\kern-.125emX}}

\usepackage[linesnumbered,commentsnumbered,ruled,vlined]{algorithm2e}
\def\BibTeX{{\rm B\kern-.05em{\sc i\kern-.025em b}\kern-.08em
    T\kern-.1667em\lower.7ex\hbox{E}\kern-.125emX}}

\begin{document}


\title{Impact of Power Consumption Models on the Energy Efficiency of Downlink NOMA Systems 


}

\author{\IEEEauthorblockN{Syllas R. C. Magalh\~{a}es, Suzan Bayhan, Geert Heijenk}
\IEEEauthorblockA{
\\University of Twente, The Netherlands\\
\{s.magalhaes, s.bayhan, geert.heijenk\}@utwente.nl}
}

\maketitle

\vspace{-35pt}

\begin{abstract}
While non-orthogonal multiple access~(NOMA) improves spectral efficiency, it results in a complexity at the receivers due to successive interference cancellation~(SIC). 
Prior studies on the energy efficiency of NOMA overlook the SIC overhead and rely on simplistic power consumption models~(PCM). 
To fill this gap, we first introduce PCM-$\kappa$ that accounts for SIC-related power expenditure, where  $\kappa$ represents the average power consumption per SIC layer. Then, to investigate the energy efficiency of NOMA and joint transmission (JT)-CoMP NOMA, 
we formulate a power allocation problem for maximizing the energy efficiency and consequently propose  a global approach running at a centralized entity and a local algorithm running at a base station. We evaluate the energy efficiency using  PCM-$\kappa$ and two PCMs commonly used in the literature. {Numerical analysis suggests that using simplistic PCMs leads to a few orders of magnitude overestimation of energy efficiency, especially when the receivers have low rate requirements. 
Despite the superiority of JT-CoMP NOMA over conventional NOMA in finding a feasible power allocation, the difference in their energy efficiency is only marginal.} 
Moreover, when 
conventional NOMA is feasible, 
the optimal solution for  JT-CoMP NOMA 
converges to conventional NOMA and NOMA schemes favour the users with the best channel quality.


\end{abstract}


\section{Introduction}

\begin{figure}[t]
 \centering
 \includegraphics[width=0.45\textwidth]{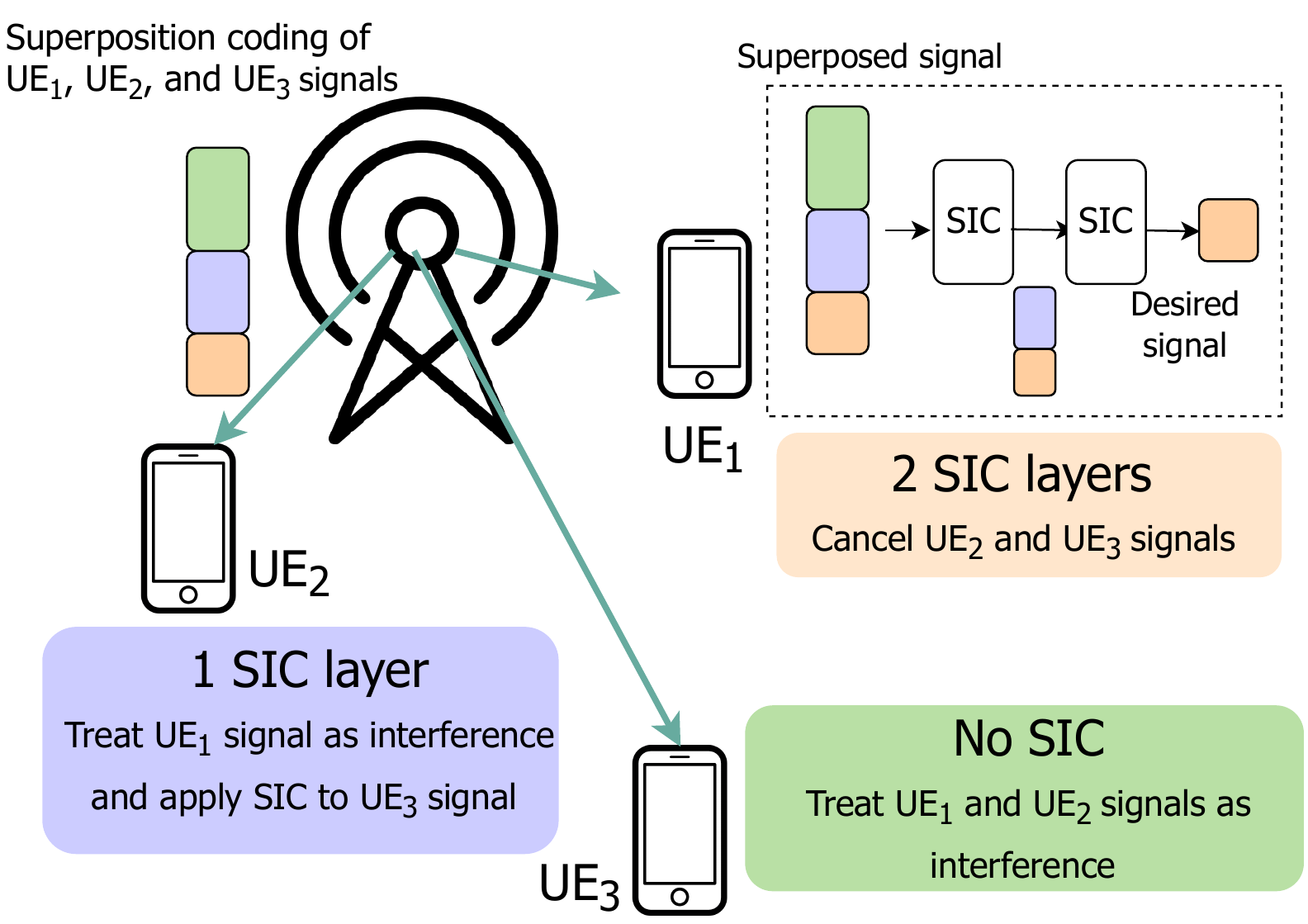}
  \caption{The NOMA transmitter applies superposition coding to the signals aimed for the users in the same NOMA cluster: UE$_1$, UE$_2$, and UE$_3$ in the above example. The NOMA receivers apply SIC to the signals aimed for the users who have a weaker channel quality and treat the other signals as interference. For instance, UE$_2$ cancels the signal for UE$_3$ who has the poorest channel quality and treats UE$_1$'s signal as interference. \label{fig:noma-toy-example} \vspace{-20pt}}
 \end{figure}

 Attaining higher spectral efficiency  has typically been the primary goal in the design of new cellular technologies to meet the increasing demand for high data-rate applications. However, with growing concerns on the energy consumption of the communication systems, 
  attaining higher energy efficiency~(EE) has also become essential~\cite{lopez2022survey}. 
A promising solution for increasing spectral efficiency is non-orthogonal multiple access~(NOMA), which suggests using the same resources to accommodate multiple users by exploiting the channel quality 
and/or data rate requirement
differences among the users and successive interference cancellation~(SIC) at the receivers. While NOMA can be implemented both in power-domain and code-domain, power-domain strategies are considered more promising due to the code-domain NOMA requiring high transmission bandwidth~\cite{aldababsa2018tutorial}. In power-domain NOMA, the transmitter allocates different powers via superposition coding to different users in the same \textit{NOMA cluster}. After receiving the superimposed signal, each receiver performs SIC to retrieve its own signal~(as shown in Fig.\,\ref{fig:noma-toy-example}). However, SIC introduces a challenge for NOMA. 
First, SIC is only possible when signals for different users can be decoded in all the users with higher decoding order 
\cite{shin2017non, rezvani2021optimal}. Moreover, for 
NOMA to yield a reasonable performance gain over orthogonal multiple access schemes, channel states of the users must differ sufficiently~\cite{ding2020unveiling}. 
Second, depending on the
decoding order
of the respective receiver relative to the other users in the same cluster, there will be different processing overhead at the receiver~\cite{tse2005fundamentals}. As an example, UE$_1$ in Fig.\,\ref{fig:noma-toy-example} with the highest channel quality has to apply two layers of SIC to retrieve its own signal in case of a channel-quality-based decoding order.\footnote{Studies such as \cite{ding2020unveiling} discuss what the SIC decoding order should be, e.g., based on channel quality or quality-of-service requirements of the receivers. }
%

Regarding the first aforementioned challenge, typically the channel-to-noise ratio (CNR)-based decoding order is used due to its optimality for single-cell sum-rate maximizing problems~\cite{yang2017on, khan2019joint}. Moreover, under certain conditions, it is also optimal for multi-cell NOMA~\cite{rezvani2021optimal}. With such decoding order, users with higher CNR need to decode the signals of the receivers with lower CNR in the same NOMA cluster. 
Consequently, cell-edge users with poor channel conditions will, most likely, not perform SIC but will treat the signals intended for the users with higher CNR as interference. Additionally, cell-edge users might be more prone to inter-cell interference (ICI)~\cite{ali2018coordinated} from the neighboring BSs. To mitigate the ICI, two or more BSs can coordinate their transmission to cell-edge users using coordinated multipoint~(CoMP) joint-transmission approach. 
Regarding the second challenge, this increased complexity at the receivers raises a question on the energy efficiency  of NOMA, especially considering an increasing number of resource-constrained end-devices.
 %



%
Although there is substantial research on the spectral efficiency of NOMA and CoMP \cite{ali2018coordinated,elhattab2020comp,hedayati2020comp}, only a few studies such as
\cite{muhammed2021resource,ali2018downlink, Liu2017Power} investigate the energy efficiency of these techniques. But, none of these studies account for energy consumption due to SIC and mostly rely on simplistic power consumption models~(PCM), e.g., total transmitted power at the BS. 
%
While the literature acknowledges the resulting overhead of SIC at the receiver, none investigates how this overhead might affect the network and the individual user EE. In this study, our goal is to fill this gap by investigating the EE of a multicell downlink NOMA and joint transmission CoMP NOMA~(JTCN) for different PCMs and to observe if and how conclusions might be affected by these PCMs. Toward this goal, we raise the following research questions:
\begin{itemize}
    \item Which PCMs are commonly used in the NOMA literature? 
    \item How does the used PCM affect the observed performance of a NOMA scheme, in particular of JTCN? 
    \item Does JTCN bring benefits over conventional NOMA considering EE when the SIC energy consumption overhead is also accounted for?
\end{itemize}

\noindent\textbf{Key contributions:} While addressing these questions, our key contributions are threefold:
\begin{itemize}[leftmargin=*]
    \item We formulate a power allocation optimization problem considering an EE-maximization objective for a two-cell downlink JTCN with minimum user rate requirements.
    \item After providing a literature review on the main PCMs for NOMA, we propose PCM-$\kappa$ that accounts for the additional processing due to the SIC at NOMA receivers depending on the number of SIC layers to be performed at the receiver.
    \item We solve the formulated problems using both distributed and centralized approaches leveraging Dinkelbach's algorithm to convert the original non-concave fractional problem into a convex problem. We assess the performance of these proposals under different PCMs to investigate the effect of PCMs on the energy efficiency and throughput performance of JTCN. Moreover, we compare JTCN and conventional NOMA with varying rate requirements of the users and under various levels of SIC overhead, i.e., various $\kappa$ values.  
    
\end{itemize}
\noindent\textbf{Key take-aways:} We have three key observations. \textit{First}, our numerical results demonstrate that there is a notable difference in terms of EE for different PCMs.
    Such disparity may result in misleading conclusions; most notably orders of magnitude overestimation of the maintained network energy efficiency, which might result in too optimistic device and network lifetimes if simpler PCMs are used. 
    \textit{Second}, our numerical analysis suggests that the key benefit of JTCN is that it can find feasible power allocation solutions while conventional NOMA cannot, due to the interference on the cell-edge user from the neighboring BSs. This is reflected as a lower outage ratio maintained by JTCN compared to conventional NOMA. 
We do not observe a statistically-significant higher energy efficiency or throughput. Surprisingly, for the majority of the scenarios, JTCN converges to conventional NOMA: the optimal operation mode is to serve the cell-edge user by one of the BSs, not jointly.  \textit{Third}, NOMA schemes should incorporate a fairness notion among the receivers' throughput or energy efficiency to prevent significant performance difference among the users in the same NOMA cluster.

\noindent\textbf{Paper's organization:} 
Section~\ref{sec:related} presents an overview of the most relevant work on EE of NOMA and JTCN. 
Section \ref{sec:sys_model} presents the system model and assumptions for each of the analysed scenarios. Next, Section~\ref{sec:pcms} provides the literature on PCMs commonly-used in the prior work. We also introduce our proposal for capturing the energy consumption due to SIC at the NOMA receivers. Section~\ref{sec:prob_form} introduces the problem formulation and proposed solutions for the formulated problem. Section~\ref{sec:sim} presents the performance analysis and simulation results. Finally, Section~\ref{sec:disc} discusses the limitations of our work while Section~\ref{sec:conc} concludes the paper.

\section{Related Work} \label{sec:related}









\begin{figure*}[t] 
    \centering
  \subfloat[Conventional NOMA.\label{fig:scenario_1}]{%
       \includegraphics[width=0.49\linewidth]{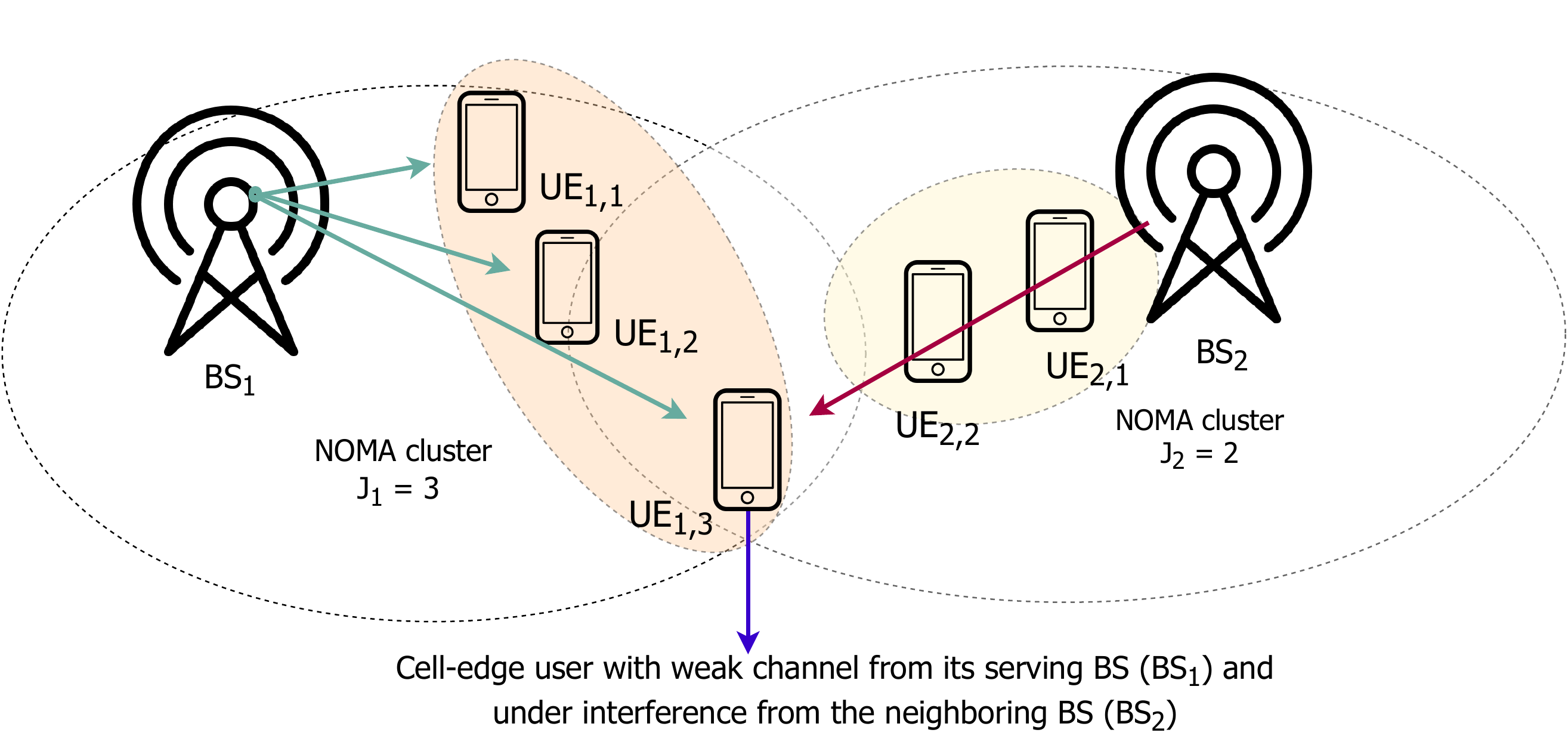}}
    \hfill
  \subfloat[Joint Transmission CoMP NOMA (JTCN).\label{fig:scenario_2}]{%
        \includegraphics[width=0.49\linewidth]{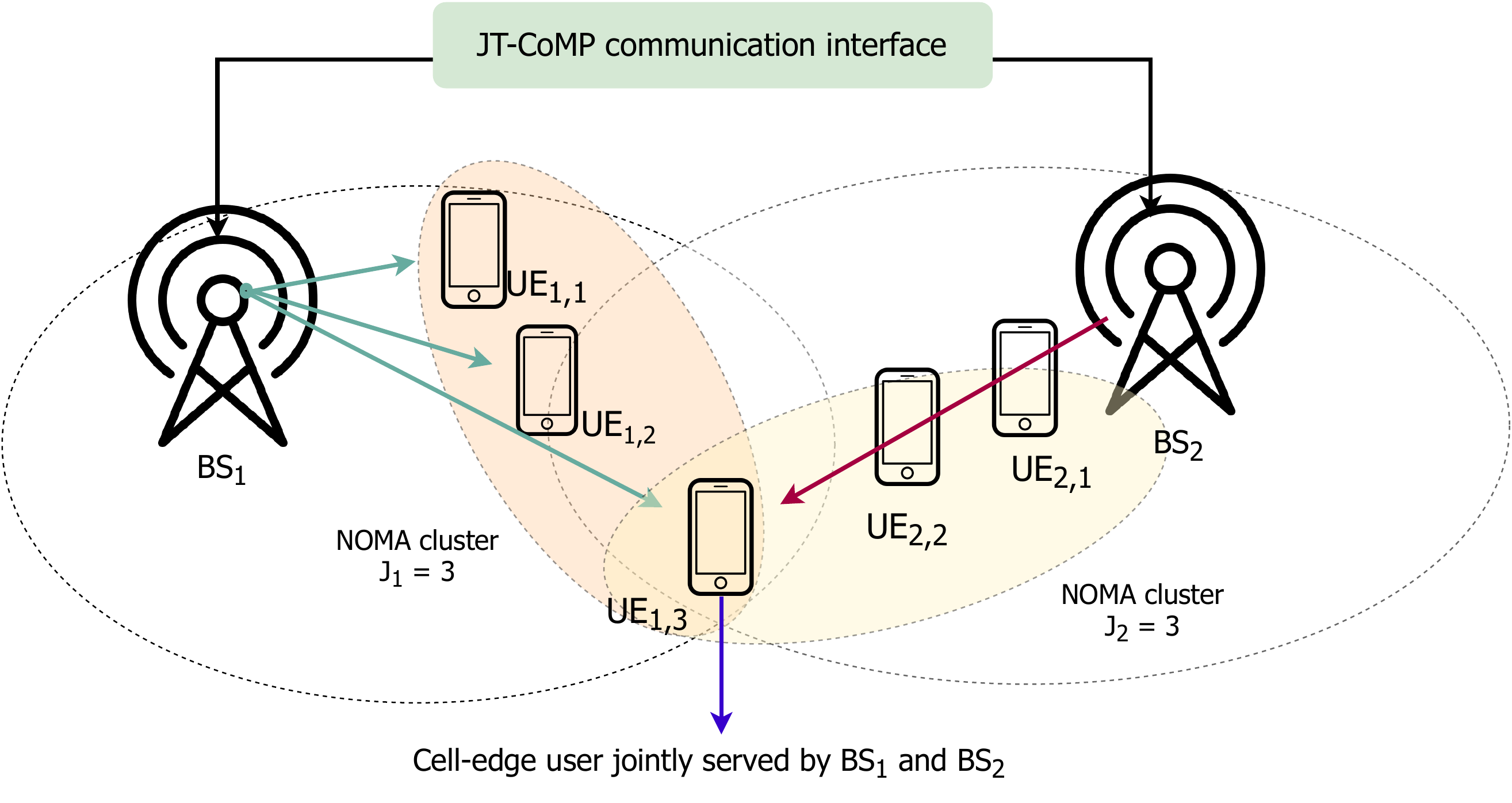}}
        \caption{Scenarios considered in the paper. a) Conventional NOMA where the cell-edge user U$_{1,3}$ is served by BS$_1$ and therefore experiences interference from BS$_2$. b) JTCN where the cell-edge user~(\textit{aka CoMP-user}) is jointly served by BS$_1$ and BS$_2$.}
  \label{fig:two-scenarios}  \vspace{-40pt}
\end{figure*}

The closest studies to ours are \cite{muhammed2021resource,ali2018downlink,Liu2017Power} which investigate the energy efficiency of NOMA and JTCN. 
Muhammed et al.\cite{muhammed2021resource} investigate the problem of EE user scheduling and power allocation for JTCN under imperfect channel state information and imperfect SIC. The authors propose a low-complexity sub-optimal solution and show that their algorithm outperforms both conventional NOMA and CoMP orthogonal frequency-division multiple access (OFDMA) in terms of EE. A similar JTCN scenario is adopted in \cite{ali2018downlink} for two-tier heterogeneous network consisting of a high-power macrocell underlaid with multiple low-power small cells with NOMA-based resource allocation. The authors formulate a sum-rate power allocation problem and propose two solutions using both a joint and a distributed approach. Besides, they provide some insights on the EE of the proposed solutions compared to the joint-transmission orthogonal multiple access (OMA) scenario. Although their simulation results show significant spectral efficiency gain, different from \cite{muhammed2021resource}, their proposed solution does not provide a significant increase in terms of EE. Another noteworthy difference between \cite{muhammed2021resource} and \cite{ali2018downlink} is their adopted PCM. While \cite{muhammed2021resource} considers a more accurate PCM, considering circuit and backhaul power, \cite{ali2018downlink} only accounts for the transmit power. In contrast to these studies, our research focuses on the energy efficiency of JTCN scenario under different PCMs.

In \cite{Liu2017Power}, the authors formulate a power allocation optimization problem and investigate the energy efficiency of a CoMP system under three different schemes, namely, JT-NOMA, JTe-NOMA and DS-NOMA. They consider universal frequency reuse, with JT-NOMA representing the scenario where all users in the system are jointly served in a single NOMA cluster, JTe-NOMA the scenario where only one cell-edge user is served by multiple BSs and DS-NOMA is the scenario where only the best channel quality BS is selected to transmit. Their simulation results show that all considered NOMA schemes outperform an OFDMA scheme and that DS-NOMA performs the best in terms of EE. However, different from our work, they assume in the JT scenario that CoMP users receive the same power from all coordinated BSs, which prevents the JT system from allocating zero power in all BSs except one for a CoMP user.



\section{System Model and Assumptions}\label{sec:sys_model}


We consider a multi-user NOMA downlink system in a network consisting of $N_{BS}$ cells, as shown in Fig.\,\ref{fig:two-scenarios}. All users and base stations (BSs) are assumed to operate with a single antenna configuration. In this system, 
BS$_b$ serves $J_b$ 
users\footnote{Note that $\sum_{b=1}^{N_{BS}} J_b$ might differ from the number of users in the system, since one user can be served by more than one BS. For example, in Fig.\,\ref{fig:scenario_1} $J_1 = 3$ and $J_2 = 2$, but for the same setup, in Fig.\,\ref{fig:scenario_2}, we have $J_1 = J_2  = 3$.}, which are at  distances $d_{1,b}, d_{2,b}, \ldots, d_{J_b,b}$ from BS$_b$, where $1\leq b \leq N_{BS}$ and $d_{1,b} \leqslant d_{2,b}, \ldots \leqslant d_{J_b,b}$.
We will denote by $u_{i,b}$ (where $0\leq i \leq J_b$) the $i^{\textrm{th}}$ user of BS$_b$. 
Moreover, we assume that there is a user located at  a random position within the coverage area of $N_{BS}$ BSs. We will refer to this user as the 
\textit{cell-edge user}. We assume that this user will be served by all $N_{BS}$ BSs in case of JTCN and by BS$_1$ in case of conventional NOMA. Additionally, we assume that there is a single NOMA cluster in each cell. 

At the beginning of a time slot, each BS determines the power allocation for its users in a NOMA cluster to ensure that each user maintains a minimum bit rate asserted by its application~($R^{\textrm{min}}_{i,b}$ bps for $u_{i,b}$). For JTCN, the cell-edge user will be served by all BSs, therefore, $u_{J_b,b}$ for all $b$ refers to the same user. Consequently, the data rate requirement $R^{\textrm{min}}_{J_b,b}$ will be a single requirement to be jointly met by all BSs. 
Additionally, we assume that the total available bandwidth is divided into 
frequency resource blocks, each with a bandwidth of $B$\,Hz. We will denote by $\omega$ the number of resource blocks assigned to each NOMA cluster. 
We assume that the frequency reuse factor is 1, i.e., all BSs transmit over the same resource blocks, resulting in some interference with each other.    


We denote by $p_{i,b}$ the power allocated by $\BS_b$ to $u_{i,b}$. The channel between the $\BS_{b^{'}}$ and $u_{i,b}$ is assumed to be a flat Rayleigh fading channel over the bandwidth $B$ and denoted by $h_{i,b,b'}$ or simply $h_{i,b}$, when $b=b'$. We assume that each BS has the channel state information for its users and calculates the channel-to-noise ratio~(CNR) denoted by  $\chgain_{i,b}$.  We can calculate $\chgain_{i,b}$ as follows: $\chgain_{i,b}= \frac{|h_{i,b}|^2}{BN_0}$ where $N_0$ is the noise power spectral density. Similarly, we calculate  $\chgain_{i,b,b'}$ which is the normalized channel gain between the $\BS_{b'}$ and $u_{i,b}$. 

Without loss of generality, let us assume that the CNR follows the inverse order of the user indices and SIC decoding on each user follows the descending order of these indices: in an $J_b$-user NOMA cluster, from the received superposition-coded signal, $u_{i,b}$ needs to cancel first the signal for $u_{J_b,b}$, next the signals for $u_{J_b-1,b}, \ldots, u_{i+1,b}$ before decoding its own signal. Similarly, $u_{J_b,b}$ is the cell-edge user and will not perform any SIC decoding. As stated earlier, we are interested in the performance of NOMA and JTCN. Next, we introduce these two scenarios depicted in Fig.\,\ref{fig:two-scenarios}. 
\subsection{Conventional NOMA}
In this case, the cell-edge user is served by only one of the BSs and each user is part of only one NOMA cluster~(as illustrated in Fig.\,\ref{fig:scenario_1} depicting two-cell scenario).  
Now, let us calculate the achievable downlink rate for $u_{i,b}$ denoted by $R^{\textrm{NOMA}}_{i,b}$.
This user will experience two types of interference: (i) \textit{inter-cell interference} from all the BSs other than its serving BS, (ii) \textit{intra-cluster interference} from the transmissions intended for the users in the same NOMA cluster served by $\BS_b$ and whose channel gain is higher than that of $u_{i,b}$~(i.e., $[u_{1,b}, \cdots, u_{i-1,b}]$).
Considering these two interference components, $R^{\textrm{NOMA}}_{i,b}$ can be expressed as follows:
\begin{align} 
    R^{\textrm{NOMA}}_{i,b}&= \omega B\log_2
    ( 1 {+} \dfrac{ p_{i,b}\chgain_{i,b}}{\sum\limits_{\substack{b'=1,\\ b'\neq b}}^{N_{BS}}\sum\limits_{j=1}^{J_{b'}} p_{j,b'}\chgain_{i,b,b'} {+} \sum\limits_{j=1}^{i-1} p_{j,b}\chgain_{i,b} {+}\omega}
    ),
    \label{eq:NOMA_data_rate}
\end{align}
where the first term in the denominator represents the inter-cell interference and the second term represents the intra-cluster interference.

\subsection{Joint Transmission CoMP NOMA~(JTCN)}
In JTCN scheme, multiple cells are coordinated to simultaneously transmit the same data to a cell-edge user~(also referred to as \textit{CoMP user}) over the same set of resource blocks. That means, unlike the previous scenario, now the cell-edge user is part of all NOMA clusters. 
We assume equal number of users in all NOMA clusters as shown in Fig.\,\ref{fig:scenario_2} for the considered example with two BSs. 
However, to guarantee SIC, the decoding order for the CoMP user needs to be the same in all NOMA clusters~\cite{ali2018coordinated}. As in the NOMA scenario, the signals intended for non-CoMP users served by other BSs are always treated as interference. 

Now let us calculate the achievable downlink rate for $u_{i,b}$. Since the cell-edge user is treated differently than other users, we need to consider these two types of users. For a non-CoMP user, we can calculate $R^{\textrm{JTCN}}_{i,b}$ similar to (\ref{eq:NOMA_data_rate}) as follows: 
\begin{align}
  R^{\textrm{JTCN}}_{i,b}= \omega B\log_2
    ( 1 + \dfrac{ p_{i,b}\chgain_{i,b}}{ \sum\limits_{\substack{b'=1,\\ b'\neq b}}^{N_{BS}}\sum\limits_{j=1}^{J_{b'}} p_{j,b'}\chgain_{i,b,b'} + \sum\limits_{j=1}^{i-1} p_{j,b}\chgain_{i,b} +\omega} 
    ).
    \label{eq:NOMA_data_rate_CoMP_1}
\end{align}
Note that the only difference between (\ref{eq:NOMA_data_rate}) and (\ref{eq:NOMA_data_rate_CoMP_1}) is that, in (\ref{eq:NOMA_data_rate_CoMP_1}), $u_{J_{b'},b'}$ refers to the same user (CoMP user) for all $b'$, $1 \leq b' \leq N_{BS}$, while in (\ref{eq:NOMA_data_rate}), $u_{J_{b'},b'}$ refers to different users for all $b$ as there are no users who are part of more than one NOMA cluster in the conventional scenario.


For the CoMP user, we express the downlink rate as follows:
\begin{align}
   R^{\textrm{JTCN}} = \omega B\log_2\left( 1 +  \dfrac{ \sum\limits_{b=1}^{N_{BS}}p_{J_{b},b}\chgain_{J_{b},b} }{ \sum\limits_{b=1}^{N_{BS}}\sum\limits_{j=1}^{J_{b} - 1} p_{j,b}\chgain_{J_{b},b} + \omega} \right).
    \label{eq:NOMA_data_rate_CoMP_2}
\end{align}
Note that different from (\ref{eq:NOMA_data_rate_CoMP_1}), the useful power  in (\ref{eq:NOMA_data_rate_CoMP_2}) is the sum of the power received from all BSs and there is only intra-cluster interference, i.e., interference from the transmissions intended for the users in the same NOMA clusters.

\section{Power Consumption Models and Our Proposal} \label{sec:pcms} 
To understand the EE of wireless networks and to design accurate EE resource allocation methods, it is crucial to have 
PCMs that account for key power consumption components. After a literature review~\cite{ali2018downlink, khan2019joint, xiao2018an,ruby2019enhanced, zhang2016energy, zhang2018energy,zamani2020optimizing, wang2013optimum}, we identify the following three PCMs 
summarized in Table~\ref{tab:literature_review}.


\begin{table}[t]
\caption{Summary of the PCMs used in the NOMA literature.}
\label{tab:literature_review}
\begin{tabularx}{\columnwidth}{|c |c| X|}
 \hline
     PCM & Papers & Description \\
 \hline
  PCM-1&    \cite{ali2018downlink, khan2019joint, xiao2018an}  & Transmitted power \\ \hline
  PCM-2 &     \cite{ruby2019enhanced, zhang2016energy, zhang2018energy} & Transmitted power and fixed circuit power\\ \hline 
PCM-3 &  \cite{zamani2020optimizing, wang2013optimum} & Transmitted, circuit, and receivers' processing power \\ \hline
\end{tabularx}
\end{table}

\begin{itemize}[leftmargin=*]
    \item PCM-1: The simplest model for power consumption in a mobile communication system takes into account only the radiated power and can be expressed as:
\begin{equation}
    P(\mathbf{p}) = \sum_{b=1}^{N_{BS}}\sum_{i=1}^{J_b}p_{i,b},
\end{equation}
where 
$\mathbf{p} = [\ldots,p_{i,b}, \ldots]$
is a vector of radiated powers with $p_{i,b}$ being the power allocated for $u_{i,b}$ in the downlink or the transmit power of $u_{i,b}$ when an uplink scenario is considered. 
As Table~\ref{tab:literature_review} shows, the following studies used this model: \cite{ali2018downlink,khan2019joint, xiao2018an}. 
\smallskip
\item{PCM-2:}
The most common way of modelling energy consumption in a cellular system is as follows:
\begin{equation}
    P(\mathbf{p}) = \sum_{b=1}^{N_{BS}} \left(\pfix + \sum_{i=1}^{J_b}p_{i,b}\right) \label{eq;pcm-2}
\end{equation}
where $\pfix$ accounts for the circuit power consumption of $\BS_b$, which is independent of the BS radiated power and includes the energy consumed by the circuitry and processing during  transmission. Studies using this model are 
\cite{ruby2019enhanced, zhang2016energy, zhang2018energy}.
\smallskip
\item{PCM-3:}
A more precise power consumption model can be expressed as: 
\begin{equation}
     P(\mathbf{p})  = \sum_{b=1}^{N_{BS}} \left(\pfix + \sum_{i=1}^{J_b}\bar{p}_{i,b}\right), \label{eq;pcm-3}
\end{equation}
where $\pfix$, as in PCM-2, is a fixed power required by $\BS_b$ for operations such as cooling, power amplifier, power supply and control signalling. In (\ref{eq;pcm-3}), $\bar{p}_{i,b}$ 
represents not only the transmission power for each user but also power consumed for signal processing for each user. This includes user-related power expenditure in both transmitter and receiver, e.g., power consumed for compression, channel coding, and modulation.
Hence, $\bar{p}_{i,b}$ is defined as follows:
\begin{align}
\bar{p}_{i,b}= p_{i,b} + p^{\text{sig}}_{i,b}(\cdot),
\end{align}
where  $p^{\text{sig}}_{i,b}(\cdot)$ can be as complex as needed to model the system power consumption accurately.




\smallskip

Our literature review reveals that a common approach is to model $p^{\text{sig}}_{i,b}(\cdot)$ as a rate-dependent power \cite{tervo2017energy, wang2013optimum, xiong2016energy, bjornson2018how}. For instance, \cite{wang2013optimum} considers $p^{\text{sig}}_{i,b}(R_{i,b}) = \rho R_{i,b}$ which is a simple example of a rate-dependent circuit power where $R_{i,b}$ is $u_{i,b}$'s data rate and $\rho>0$ is a constant value. Another example is ~\cite{zamani2020optimizing} which assumes for a single-cell NOMA scenario a linear function of the allocated transmission power, i.e., $p^{\text{sig}}_{i,b}(p_{i,b}) = \rho p_{i,b}$. While the PCM used in \cite{zamani2020optimizing} is more realistic and is a generic model representing also PCM-1 and PCM-2 as its special cases, it neglects the differences among the NOMA receivers while they decode their signals. 
As discussed earlier, NOMA requires SIC at the receiver side and the SIC overhead depends on the index~(decoding order) of a user in its NOMA cluster. For instance, the cell-edge user with the weakest channel does not apply SIC as there is no other user with a weaker channel. On the other hand, the user with the best channel quality might need to run many layers of SIC depending on the size of the NOMA cluster. 
Therefore, in our power consumption model, we need to consider a NOMA user's rank in its cluster when users are sorted according to their CNRs. To the best of our knowledge, no prior work models this difference among NOMA receivers. 
\end{itemize}

\smallskip
\noindent
\textbf{{Proposed PCM considering SIC~(PCM-$\kappa$):}} To account for additional processing performed at each NOMA receiver, we propose PCM-$\kappa$ based on the number of decodings at each user. Let us denote by $\kappa$ the average power expenditure for each decoding during the SIC processing at the receiver. In practice, $\kappa$ depends on the bitrate of each superposed signal decoded in a receiver applying SIC~\cite{mezghani2011power, bjornson2015optimal}. However, for the sake of simplicity, 
we assume a fixed average power consumption for performing one SIC layer. {As~\cite{zamani2020optimizing}, our model also reflects the dependence of the coding/decoding power consumption on the transmission power allocated to a NOMA receiver.}  
%
More formally, we define power consumption at $u_{i,b}$ as follows:
\begin{equation}
    p_{sig,i}(\mathbf{p}) = \rho p_{i,b} + (J_b - i + 1)\kappa,
\end{equation}
where 
$(J_b - i + 1)$ accounts for the number of decodings necessary at $u_{i,b}$. The power consumption in the network is then formulated as follows:
\begin{align} 
 P(\mathbf{p})  = \sum_{b=1}^{N_{BS}} \left( \pfix {+} \sum_{i=1}^{J_b} \left((1+\rho)p_{i,b}+(J_b - i + 1)\kappa\right)\right). \label{eq;pcm-proposed} 
\end{align}

Please note that our model aims at incorporating the overhead due to SIC as it is known to be a complex process and many studies therefore assumed only very small NOMA clusters, e.g., only two users in  \cite{muhammed2019energy}. However, our model needs to be validated in a testbed for its accuracy, which we leave as a future work. 
%


\vspace{-20pt}
\section{Optimal Power Allocation Problems}
\label{sec:prob_form}
This section first presents our problem formulation for EE-maximization for the considered NOMA scenarios in Sec.\,\ref{sec:prob_form_noma} and Sec.\,\ref{sec:prob_form_jt}.  
%
Due to the complexity of solving these formulated problems optimally, in Sec.\,\ref{sec:solutions}, we present a series of transformations so that Dinkelbach's algorithm, which is typically used for solving fractional programs, can be applied to our problem.  

\vspace{-20pt}
\subsection{Conventional NOMA Problem Formulation}\label{sec:prob_form_noma}



Let us denote by $\mathbf{p}$ our decision variable, which is the power allocation vector $[\ldots,  p_{i,b}, \ldots]$.
We will denote by $P^{}(\mathbf{p})$ the corresponding power consumption in the system according to the considered PCM.  Then, we can formulate the EE-maximizing power allocation problem as follows:
\begin{align}
&\textrm{P1:} \max_{\mathbf{p}} \dfrac{\sum\limits_{b=1}^{N_{BS}}\sum\limits_{i=1}^{J_b} R^{\textrm{NOMA}}_{i,b}}{P^{}(\mathbf{p})}\label{eq:conv_obj_func_EE}\\
&\textrm{subject to:}\nonumber\\
&R^{\textrm{NOMA}}_{i,b} \geq R^{\textrm{min}}_{i,b}, \ \substack{ b = 1,\ldots,N_{BS} \\ i = 1,\ldots,J_b }\label{eq:conv_problem_rate_constraint}\\ &\sum\limits_{i=1}^{J_b}p_{i,b} \leq P_{max},\ b = 1,\ldots,N_{BS} \label{eq:conv_problem_maxP_constraint}\\
&p_{i,b} \geq 0, \ \substack{ b = 1,\ldots,N_{BS} \\ i = 1,\ldots,J_b} \label{eq:conv_problem_nonneg_constraint}
\end{align}
where the numerator in (\ref{eq:conv_obj_func_EE}) accounts for the sum of the received data rate of all users. Additionally, 
Const. (\ref{eq:conv_problem_rate_constraint}) ensures that $u_{i,b}$ 
will maintain a data rate of at least $R_{i,b}$ bps.  Const. (\ref{eq:conv_problem_maxP_constraint}) guarantees that the BS radiated power is equal or lower than the available transmission power 
while Const. (\ref{eq:conv_problem_nonneg_constraint}) ensures that assigned power values are non-negative.

\subsection{JTCN Problem Formulation}\label{sec:prob_form_jt}

The EE-maximizing power allocation problem for JTCN scenario can be formulated as follows:
\begin{align}
&\textrm{P2:}  \max_{\mathbf{p}} \dfrac{ R^{\textrm{JTCN}} + \sum\limits_{b=1}^{N_{BS}}\sum\limits_{i=1}^{J_b - 1} R_{i,b}^{\textrm{JTCN}}}{P^{}(\mathbf{p})}\label{eq:JT_obj_func_EE} \\
&\textrm{subject to:} \nonumber \\
&R^{\textrm{JTCN}}_{i,b} \geq R^{\textrm{min}}_{i,b}, \ \substack{b = 1,\ldots,N_{BS} \\ i = 1,\ldots,J_b - 1}\label{eq:JT_problem_rate_constraint} \\
&R^{\textrm{JTCN}} \geq R^{\textrm{min}} \label{eq:JT_problem_rate_constraint_CoMP_user} \\
& \sum\limits_{i=1}^{J_b}p_{i,b} \leq P_{max},\ b = 1,\ldots,N_{BS} \label{eq:JT_problem_maxP_constraint} \\
&p_{i,b} \geq 0, \ \substack{b = 1,\ldots,N_{BS} \\ i = 1,\ldots,J_b} \label{eq:JT_problem_nonneg_constraint}
\end{align}
where Const.\,(\ref{eq:JT_problem_maxP_constraint}) ensures a limit in the radiated power per BS and (\ref{eq:JT_problem_nonneg_constraint}) guarantees that assigned power values are non-negative. The minimum rate requirement, on the other hand, is divided into two constraints, namely Const.\,(\ref{eq:JT_problem_rate_constraint}) and Const.\,(\ref{eq:JT_problem_rate_constraint_CoMP_user}) in which the former ensures that each non-CoMP user will achieve at least $R^{\textrm{min}}_{i,b}$ bps and the latter guarantees that the CoMP user's rate is at least $R^{\textrm{min}}$ bps.


%
\subsection{Solution Approaches}\label{sec:solutions}

\begin{figure*}[t]
\subfloat[Global approach running at a central node.]{\includegraphics[width=0.49\textwidth]{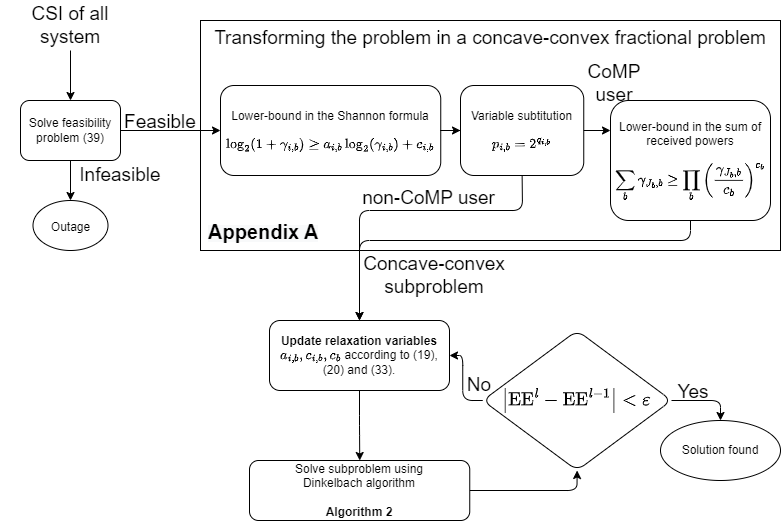}
\label{fig:flowchart_global}}%
\subfloat[ILO running at each BS.]
{\includegraphics[width=0.49\textwidth]{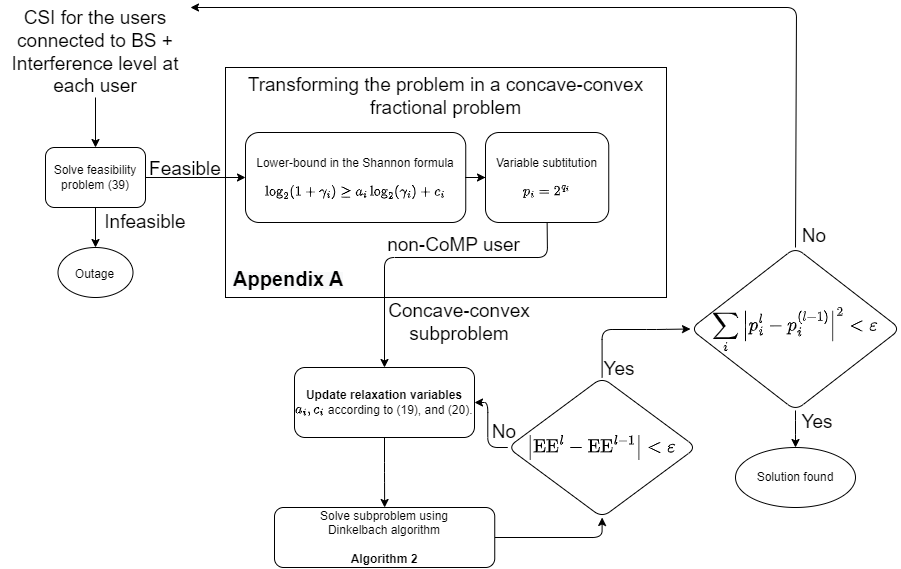}
\label{fig:flowchart_local}}
\caption{Summary of the solution approaches.}%
\label{fig:flowcharts} 
\vspace{-0.5cm}
\end{figure*} 

To solve the formulated problems, we propose the following two approaches: (i) \textit{global} and (ii) \textit{iterative local optimization}~(ILO). The first approach assumes a central node with channel state information of all channels between users and BSs where the optimization can be performed globally, i.e., directly solving 
P1 and P2.
ILO, on the other hand, uses an iterative algorithm to solve the problem locally at each BS. 
Next, we present these two approaches whose steps are summarized in Fig.\,\ref{fig:flowchart_global} and Fig.\,\ref{fig:flowchart_local}.

\subsubsection{Global Optimization (summarized in Alg.\ref{alg:global})} 
Solving the EE problems
P1 and P2 
is challenging because of the fractional form of the objective functions, which leads to non-concave maximization problems. Moreover, due to the particularity of the CoMP user data rate expression, the solution approach for 
P1 and P2 
differs slightly. Therefore, we will first focus on the conventional scenario.

To solve 
P1, besides the convexity of the constraints, we need a concave numerator and a convex denominator in the transmit powers, so that Dinkelbach's algorithm can be applied \cite{dinkelbach1967nonlinear}. Regardless of the adopted PCM, the denominator is affine so is convex. The numerator, on the other hand, is not concave since it is the sum of a difference of convex functions (it can be seen by making the expression inside the logarithm a single fraction and then making it a difference of logarithms), which is not concave \cite{boyd2004convex}. To make it concave, 
we apply a lower-bound and a variable change in the original expression (see Appendix~\ref{ap:A}) to obtain:

\begin{align}
    R_{i,b}^{\textrm{NOMA}} &\geq a_{i,b}\omega B(\log_2\left(\chgain_{i,b}\right) + q_{i,b})+ c_{i,b}\omega B  - a_{i,b}\omega B\log_2\left( \sum\limits_{\substack{b'=1,\\ b'\neq b}}^{N_{BS}}\sum\limits_{j=1}^{J_{b'}} 2^{q_{j,b'}}\chgain_{i,b,b'} + \sum\limits_{j=1}^{i-1} 2^{q_{j,b}}\chgain_{i,b} +\omega \right)\nonumber\\
    &=\Tilde{R}_{i,b}^{\textrm{NOMA}},
   \nonumber
\end{align}
where $p_{i,b} = 2^{q_{i,b}}$ and the equality holds when $\{a_{i,b}, c_{i,b}\}_{\forall i, \forall b}$ are computed using (\ref{eq:a}) and (\ref{eq:c}).

Therefore, the objective function can now be written as a fraction with a concave numerator and a convex denominator and we can rewrite P1 as:
\begin{align}
&\max_{\mathbf{q}} \dfrac{\sum\limits_{b=1}^{N_{BS}}\sum\limits_{i=1}^{J_b} \Tilde{R}_{i,b}^{\textrm{NOMA}} }{P^{}|_{p_{i,b} = 2^{q_{i,b}}}}\label{eq:conv_obj_func_EE_2} \\
&\textrm{s.t.}
\left.R_{i,b}^{\textrm{NOMA}}\right|_{p_{i,b} = 2^{q_{i,b}}} \geq R^{\textrm{min}}_{i,b},
\substack{b = 1,\ldots,N_{BS} \\ i = 1,\ldots,J_b}\label{eq:conv_problem_rate_constraint_2} \\
 &\sum\limits_{i=1}^{J}2^{q_{i,b}} \leq P_{max},\ b = 1,\ldots,N_{BS}. \label{eq:conv_problem_maxP_constraint_2}
\end{align}

To apply the Dinkelbach's algorithm, all constraints must be convex, which is not the case for (\ref{eq:conv_problem_rate_constraint_2}) with the new optimization variable. To make it convex, let us first define $\gamma^{min}_{i,b} = 2^{\frac{R^{\textrm{min}}_{i,b}}{\omega B}}-1$. We can, then, manipulate (\ref{eq:conv_problem_rate_constraint_2}) as follows:
\begin{align} 
    2^{\dfrac{\left.R_{i,b}^{\textrm{NOMA}}\right|_{p_{i,b} = 2^{q_{i,b}}}}{\omega B}} - 1 \geq 2^{\dfrac{R^{\textrm{min}}_{i,b}}{\omega B}} -1,
\end{align}
i.e.,
\begin{align} 
    \dfrac{ 2^{q_{i,b}}\chgain_{i,b}}{ \sum\limits_{\substack{b'=1,\\ b'\neq b}}^{N_{BS}}\sum\limits_{j=1}^{J_{b'}} 2^{q_{j,b'}}\chgain_{i,b,b'} + \sum\limits_{j=1}^{i-1} 2^{q_{j,b}}\chgain_{i,b} +\omega} \geq \gamma^{min}_{i,b}.
    \label{eq:conv_problem_rate_constraint_3_step_2}
\end{align}
For $R_{i,b} > 0$, we have $\gamma^{min}_{i,b} > 0$, which allows us to divide both sides of (\ref{eq:conv_problem_rate_constraint_3_step_2}) by $\gamma^{min}_{i,b}$. Additionally, multiplying by $-1$ and applying $\log_2(\cdot)$ on both sides of (\ref{eq:conv_problem_rate_constraint_3_step_2}) leads to:
\begin{align} 
    -q_{i,b} + \log_2\left(\sum\limits_{\substack{b'=1,\\ b'\neq b}}^{N_{BS}}\sum\limits_{j=1}^{J_{b'}} 2^{q_{j,b'}}\chgain_{i,b,b'} + \sum\limits_{j=1}^{i-1} 2^{q_{j,b}}\chgain_{i,b} +\omega\right) - \log_2\left(\dfrac{\chgain_{i,b}}{\gamma^{min}_{i,b}}\right) \leq 0.
    \label{eq:conv_problem_rate_constraint_3}
\end{align}




Now, (\ref{eq:conv_problem_rate_constraint_3}) is convex w.r.t. $q_{i,b}$ and we can apply the Dinkelbach's algorithm with the fractional problem to be solved expressed as:
\begin{align}
&\textrm{P3:} \max_{\mathbf{q}} \dfrac{\sum\limits_{b=1}^{N_{BS}}\sum\limits_{i=1}^{J_b} \Tilde{R}_{i,b}^{\textrm{NOMA}} }{P^{}|_{p_{i,b} = 2^{q_{i,b}}}}\label{eq:conv_obj_func_EE_3}\\
&\textrm{s.t } (\ref{eq:conv_problem_rate_constraint_3}) \textrm{ and }  
\sum\limits_{i=1}^{J_b}2^{q_{i,b}} \leq P_{max},\ b = 1,\ldots,N_{BS}. \nonumber 
\end{align}


The same approach can be applied to the non-CoMP users' data rate expression for JTCN in~(\ref{eq:NOMA_data_rate_CoMP_1}) to write $R^{\textrm{JTCN}}_{i,b}$, $1 \leq i \leq J_b -1$, as: 
\begin{align}
    \Tilde{R}_{i,b}^{\textrm{JTCN}}=a_{i,b}\omega B(\log_2\left(\chgain_{i,b}\right) + q_{i,b})+ c_{i,b}\omega B - a_{i,b}\omega B\log_2\left( \sum\limits_{\substack{b'=1,\\ b'\neq b}}^{N_{BS}}\sum\limits_{j=1}^{J_{b'}} 2^{q_{j,b'}}\chgain_{i,b,b'} + \sum\limits_{j=1}^{i-1} 2^{q_{j,b}}\chgain_{i,b} +\omega \right).
    \nonumber
\end{align}
In this scenario, however, to make the CoMP user data rate expression concave, a new lower-bound is introduced~(please refer to Appendix \ref{ap:A} for more details):
\begin{align}
    &R^{\textrm{JTCN}} \geq\Tilde{R}^{\textrm{JTCN}} = a_{i,b}\omega B \sum\limits_{b'=1}^{N_{BS}}c^{(1)}_{b'}\left( q_{J_b,b'} + \log_2\left( \dfrac{\chgain_{J_b,b'}}{c^{(1)}_{b'}} \right) \right)  \nonumber\\
   &\quad- a_{i,b}\omega B \log_2\left( \sum\limits_{b'=1}^{N_{BS}}\sum\limits_{j=1}^{J_{b'} - 1} 2^{q_{j,b'}}\chgain_{J_b,b,b'} + \omega \right) + c_{i,b}\omega B. \nonumber
\end{align}

Then, P2 can be rewritten as a concave-convex fractional problem as:
\begin{align}
&\textrm{P4:} \max_{\mathbf{q}} \dfrac{\sum\limits_{b=1}^{N_{BS}}\sum\limits_{i=1}^{J_b-1} \Tilde{R}_{i,b}^{\textrm{JTCN}} + \Tilde{R}^{\textrm{JTCN}} }{P^{}|_{p_{i,b} = 2^{q_{i,b}}}}\label{eq:JT_obj_func_EE_3}\\
&\textrm{s.t: }  
(\ref{eq:JT_problem_rate_constraint_3}),(\ref{eq:conv_problem_rate_constraint_3}),  
 \textrm{ and } \sum\limits_{i=1}^{J_b} 2^{q_{i,b}} \leq P_{max},\ b = 1,\ldots,N_{BS} \nonumber
\end{align}

 \begin{algorithm} [t]
 \caption{Global optimization}
 \label{alg:global}
 \begin{algorithmic}[1]
 \STATE Solve the energy minimization problem:\\ $\min_{\mathbf{p}} \sum_{b=1}^{N_{BS}}\sum_{i=1}^{J_b} p_{i,b}$, s.t. (\ref{eq:conv_problem_rate_constraint}), (\ref{eq:conv_problem_maxP_constraint}) and (\ref{eq:conv_problem_nonneg_constraint}) for conventional NOMA, and s.t. (\ref{eq:JT_problem_rate_constraint}), (\ref{eq:JT_problem_rate_constraint_CoMP_user}),
(\ref{eq:JT_problem_maxP_constraint}) and
(\ref{eq:JT_problem_nonneg_constraint}) for JTCN;
 \IF {Feasible}
      \STATE $\epsilon >0$; $l=0$; select $\mathbf{p}^{(0)} = \mathbf{p}^*$ (the solution of the energy minimization problem);
      \WHILE {$|\text{EE}(\mathbf{p}^{(l+1)}) - \text{EE}(\mathbf{p}^{(l)})| > \epsilon$}
          \STATE $l = l + 1$;
          \STATE Compute $\{a_{i,b}^{(l)}, c_{i,b}^{(l)}, c^{(1)(l)}_{b}\}_{\forall i, \forall b}$, according to (\ref{eq:a}), (\ref{eq:c}) and (\ref{eq:c1_b});
          \STATE Compute $\mathbf{q}^{(l)}$ as the solution of problem $P^{(l)}$ ((\ref{eq:conv_obj_func_EE_3}) or (\ref{eq:JT_obj_func_EE_3})) using Algorithm~\ref{alg:2};
          \STATE $\mathbf{p}^{(l)} = 2^{\mathbf{q}^{(l)}}$;
      \ENDWHILE
  \ENDIF
 \RETURN $\mathbf{p}^{(l)}$ 
 \end{algorithmic} 
 \end{algorithm}

To guarantee the convergence of 
P3 and P4
to the solution of the respective original problems
P1 and P2, we update iteratively the coefficients $\{a_{i,b}, c_{i,b}, c^{(1)}_{b}\}_{\forall i, \forall b}$, according to (\ref{eq:a}), (\ref{eq:c}) and (\ref{eq:c1_b}). The last coefficient is  updated only in JTCN problem P4.
Please refer to Proposition 4.2 and Alg.~8 from \cite{zappone2015energy} for more details.

For solving the original problems, the objective functions of the respective subproblems P3 and P4 are first converted in a subtractive form and then iteratively solved by using the Dinkelbach's algorithm for each updated value of the coefficients $a_{i,b}$, $c_{i,b}$ and $c^{(1)}_{b}$. Defining $\Tilde{R}$ as the numerator of the objective function of the respective problem, i.e., $\Tilde{R} = \sum\limits_{b=1}^{N_{BS}}\sum\limits_{i=1}^{J_b} \Tilde{R}_{i,b}^{\textrm{NOMA}}$ for conventional NOMA and $\Tilde{R} = \sum\limits_{b=1}^{N_{BS}}\sum\limits_{i=1}^{J_b-1} \Tilde{R}_{i,b}^{\textrm{JTCN}} + \Tilde{R}^{\textrm{JTCN}}$ for JTCN, we can define a generic subproblem in the subtractive form as:
\begin{equation}
\max_{\mathbf{q}} \Tilde{R} - \lambda^{(l-1)} P^{}|_{p_{i,b} = 2^{q_{i,b}}}
\label{eq:dinkelbach_function}, 
\end{equation}
subject to the respective problem constraints.

In (\ref{eq:dinkelbach_function}), $l \in {1,2, \ldots, L_{max}}$ denotes the index of the iteration, $L_{max}$ is the maximum number of iterations and $\lambda^{(l-1)}$ is the optimal energy efficiency of the subproblem solved in the last iteration ($l-1$), given by:
\begin{equation}
    \lambda^{(l-1)} = \dfrac{ \Tilde{R}(\{q_{i',b'}^{(l-1)}\}) }{P^{}(\{q_{i',b'}^{(l-1)}\})}, \nonumber
\end{equation}
where $\{q_{i',b'}^{(l-1)}\}$ denotes the solution for the optimization subproblem at the iteration $l-1$.

Solving the original problem is equivalent to finding the maximum energy efficiency $\lambda^{*} = \frac{\Tilde{R}(\{q_{i',b'}^{*}\}) }{P^{}(\{q_{i',b'}^{*}\})}$, which can be achieved if and only if  \cite{zappone2015energy}:
\begin{equation}
     \Tilde{R}(\{q_{i',b'}^{(l)}\}) - \lambda^{*} P(\{q_{i',b'}^{(l)}\}) = 0. \nonumber
\end{equation}


 \begin{algorithm}[t]
 \caption{Dinkelbach's algorithm}
 \label{alg:2}
 \begin{algorithmic}[1]
  \STATE $\epsilon >0$; $l=0$; $\lambda^{(0)} = 0$;
  \WHILE {$F(\lambda^{(l)}) > \epsilon$}
  \STATE $\mathbf{q}^{(l)} = \arg \max_{\mathbf{q}} \left( \Tilde{R}(\mathbf{q}) - \lambda^{(l)} P(\mathbf{q}) \right)$;
  \STATE $F(\lambda^{(l)}) = \Tilde{R}(\mathbf{q}^{(l)}) - \lambda^{(l)} P(\mathbf{q}^{(l)})$;
  \STATE $\lambda^{(l+1)} = \dfrac{\Tilde{R}(\mathbf{q}^{(l)})}{P(\mathbf{q}^{(l)})}$;
  \STATE $l = l + 1$;
  \ENDWHILE
 \RETURN $\mathbf{q}^{(l)}$ 
 \end{algorithmic} 
 \end{algorithm}
 
\subsubsection{Iterative Local Optimization (ILO)}

For conventional NOMA scenario, ILO aims at finding a power allocation without requiring global knowledge and a centralized entity. In this case, we assume that the cell-edge user is served by the BS providing the best channel conditions and  channel quality remains the same during the rounds of optimization. Additionally, we assume that each BS knows the received ICI from other BSs. 
 At each BS, first, the following simpler problem is solved to find the minimum total power needed to meet the users' rate requirements:
\begin{equation}
\min_{\mathbf{p}} \sum\limits_{i=1}^{J_b} p_{i,b} \quad \textrm{where } b=1,\ldots,N_{BS}
\label{problem_min_p}
\end{equation}
subject to (\ref{eq:conv_problem_rate_constraint})-(\ref{eq:conv_problem_nonneg_constraint}). 
Then, the solution of (\ref{problem_min_p}) is fed as input to the following energy-efficiency maximization problem:
\begin{align}
&\max_{\mathbf{p}_b} \dfrac{\sum\limits_{i=1}^{J_b} R_{i,b}^{\textrm{NOMA}}}{P_b(\mathbf{p}_b)} \nonumber \\
& \textrm{s.t. } \sum\limits_{i=1}^{J_b}p_{i,b} \leq P_{max}, \label{eq:problem_constraint_1}  \nonumber \\
&p_{i,b} \geq 0, \ i = 1,\ldots,J_b \nonumber \\
&R_{i,b}^{\textrm{NOMA}} \geq R_{i,b},\ i = 1,\ldots,J_b. \nonumber 
\end{align}
The main difference between global approach and ILO  is that the former considers whole system in power allocation and the later one BS only at an iteration. However, the solution for ILO can be found following the same steps presented for global optimization. 

\section{Performance Analysis}
\label{sec:sim}
In this section, we analyze the performance of our proposals via Monte Carlo simulations with a goal of addressing the following questions: 
(i) How does the PCM used for optimizing the power allocation affect conclusions on the energy efficiency of JTCN?
(ii) Does JTCN offer benefits over NOMA in terms of energy efficiency?
(iii) How do the minimum rate requirements affect the performance? 
(iv) How does the power cost~$\kappa$ of one SIC layer decoding affect the performance? 
(v) Can a local optimization approach perform comparably to a global approach?



\begin{table}
    \centering
    \caption{Simulation parameters\label{tab:simu_param}}
    \begin{tabular}{|l|l|}
        \hline
        Parameter                   & Value    \\
        \hline
        Cell radius                 & 600m      \\
        Max. BS transmit power~($P_{max}$)     & 43 dBm \\
        Noise spectral density~($N_0$)       & -139 dBm/Hz \\
        Bandwidth of a RB ($B$)          & 180 KHz   \\Path loss model & Macrocell pathloss  \cite{3gpp2016plmodel} \\
        \# of users per NOMA cluster & 2 or 3       \\
        \# of RBs per NOMA cluster~($\omega$)   & 100 \\
        BS circuit power ($P^{\textrm{fix}}$)& 30 dBm \cite{Liu2017Power}   \\ 
        Mean power/SIC layer ($\kappa$) & \{0, 0.5, 2.5\} Watts  \\
       {Signal processing overhead} ($\rho$)     & 0.1 \\
        \hline
    \end{tabular}
\end{table}

\begin{figure*}[ht]\vspace{-10pt}
\subfloat[Optimized for PCM-1, evaluated with all PCMs.]{\includegraphics[width=0.33\linewidth]{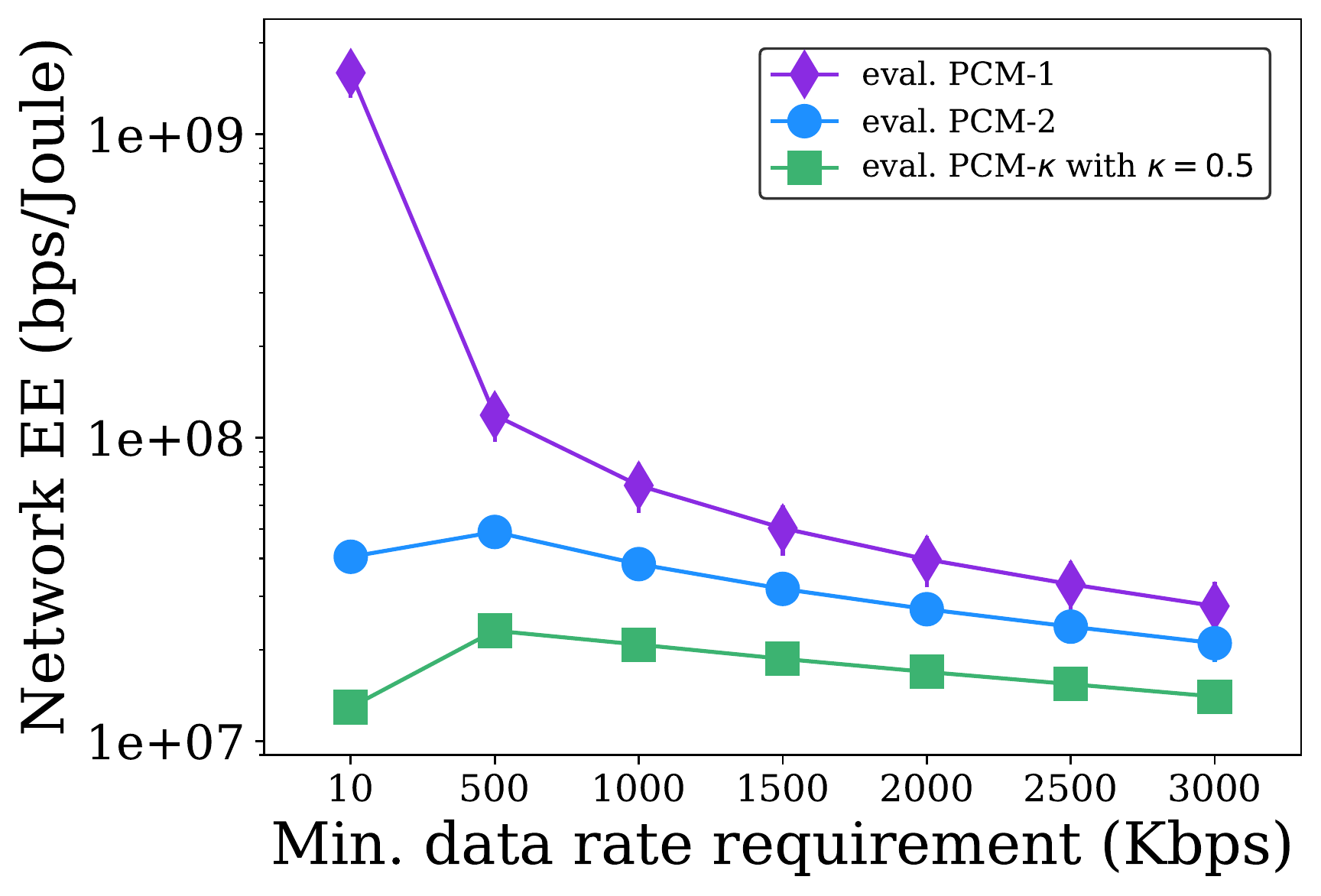}
\label{fig:EE_JT_opt_PCM1_eval_all_PCMsfig}}
\subfloat[Optimized using different PCMs.]{\includegraphics[width=0.33\linewidth]{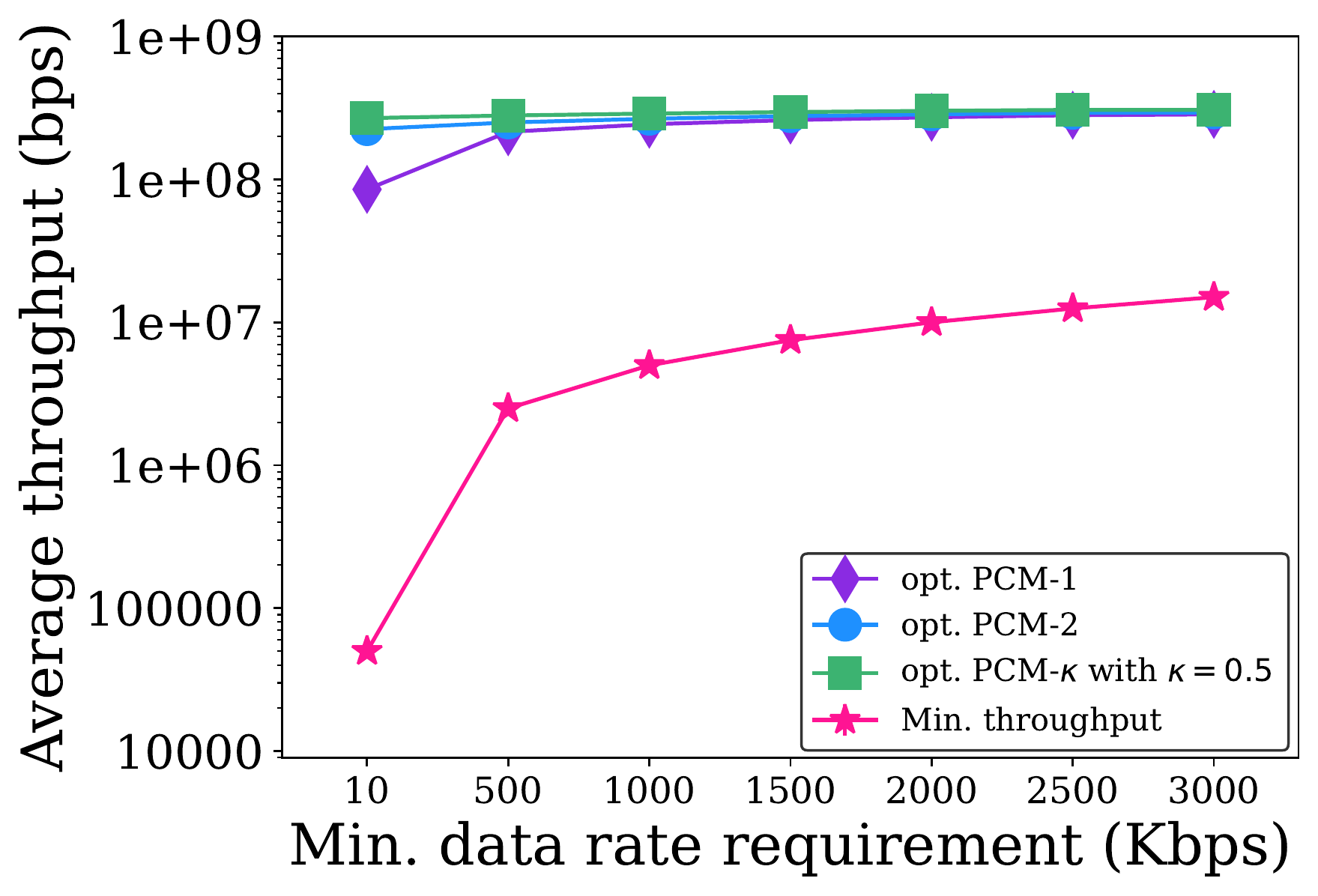}
\label{fig:TP_JT_opt_all_eval_PCMprop}}
\subfloat[Evaluated with PCM$-\kappa$.]{\includegraphics[width=0.33\linewidth]{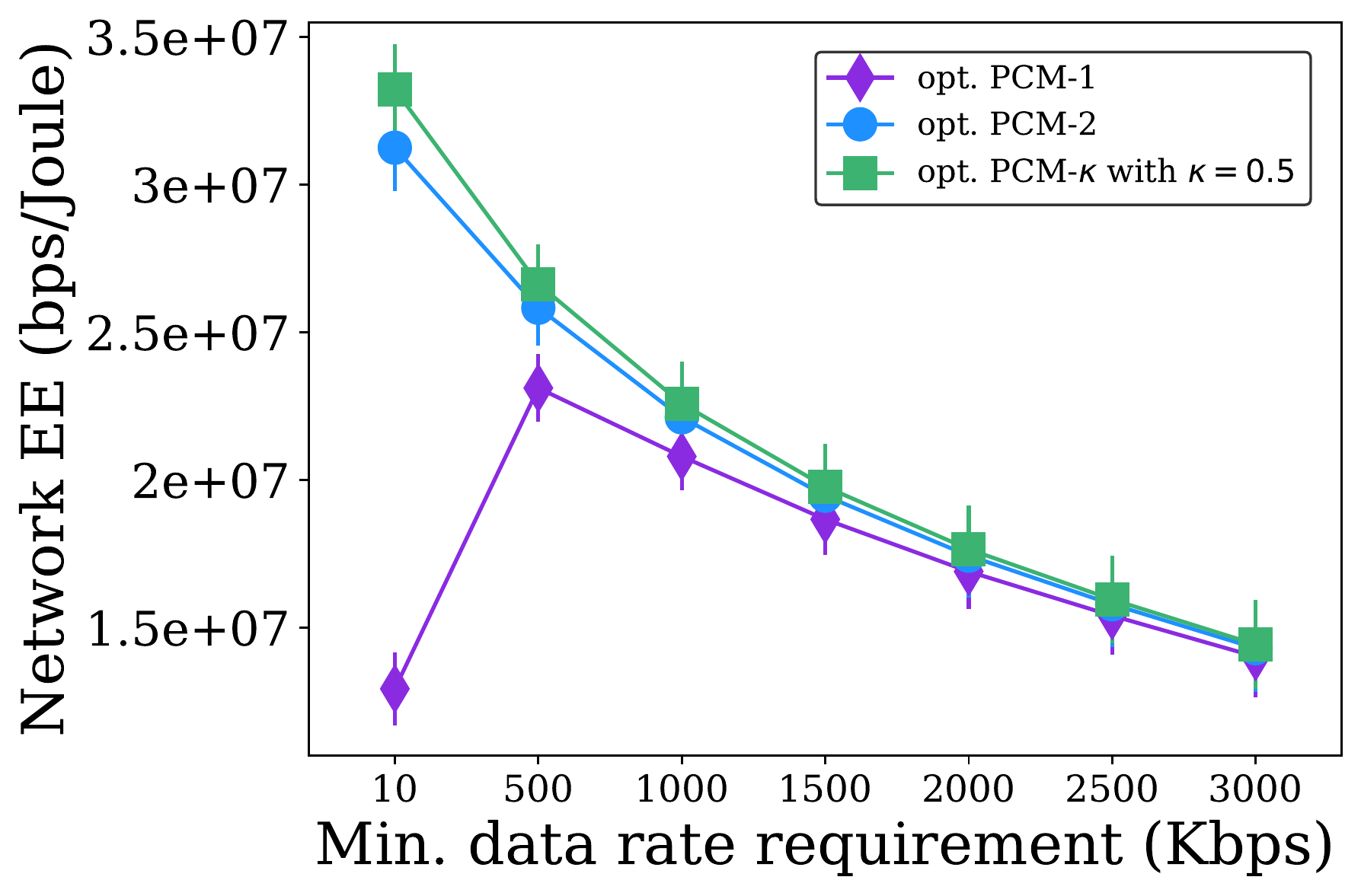}
\label{fig:EE_JT_opt_all_eval_PCMprop}}
  \caption{Energy efficiency and throughput of JTCN optimized under increasing $R^{\textrm{min}}$ considering different PCMs.} %
  \label{fig:incrate_PCMs_JTCN} \vspace{-10pt}
\end{figure*}


We assume that non-CoMP users are located from their BSs with a distance of $30$\,m and $200$\,m for all BSs. Moreover, 
the total number of available RBs for the network is $\omega = 100$ RBs, each with a bandwidth of $B = 180$\,KHz. In addition, each BS has a maximum transmission power of $P_{max} = 43$\,dBm  and the noise power spectral density is $N_0=-139$\,dBm/Hz.
The remaining simulation parameters are as follows: the radius of both BSs is $600$\,m; the distance between BS$_1$ and BS$_2$ is $1$\,km;
and path loss is modelled according to 3GPP as in \cite{3gpp2016plmodel}. We will assume that all users have the same rate requirement. Hence, we will refer to this minimum required rate by $R^{\textrm{min}}$.   Table \ref{tab:simu_param} summarizes the key  simulation parameters.

For statistical significance, we  report the following three performance metrics averaged over 100 runs with 95\% confidence intervals: network energy efficiency, network throughput, and outage ratio. 
\begin{itemize}
    \item {Network Energy Efficiency:} To measure the network energy efficiency fairly among the PCMs, we assume that the real-world power consumption is as in (\ref{eq;pcm-proposed}) which accounts for the SIC overhead. Therefore, the network energy efficiency is calculated independent of the PCM considered in the power allocation. More formally, EE is computed as follows:
\begin{equation}
     EE = \dfrac{\sum\limits_{b=1}^{N_{BS}}\sum\limits_{i=1}^{J_b} R_{i,b}}{\sum\limits_{b=1}^{N_{BS}} \left( \pfix {+} \sum\limits_{i=1}^{J_b} \{(1+\rho)p_{i,b}+(J_b - i + 1)\kappa\}\right)} \nonumber
\end{equation}  
where $R_{i,b}$ is $R^{\textrm{NOMA}}_{i,b}$ for NOMA and   $R^{\textrm{JTCN}}_{i,b}$ for JTCN. 

\item{Network Throughput:}
The throughput is defined as the sum of the achievable data rate of all users in the network and calculated as:
\begin{equation}
     \text{Throughput} = \sum\limits_{b=1}^{N_{BS}}\sum\limits_{i=1}^{J_b} R_{i,b}. \nonumber
\end{equation}

\item{Outage ratio:} When there exists at least one user whose minimum rate requirement is not met, we refer to this case as \textit{outage}. In such outage cases, the minimum required power to guarantee the rate requirements of all users is higher than the available power. Then, outage ratio is the fraction of runs with outage over all the runs, e.g., 100 MC runs. 
\end{itemize}


\subsection{Impact of the Power Consumption Model}

To understand the impact of the PCM on the network energy efficiency, we first focus on the global approach for JTCN  and analyze its performance under different PCMs both in power allocation and evaluation of the results. Fig.\,\ref{fig:EE_JT_opt_PCM1_eval_all_PCMsfig} shows the average energy efficiency under increasing $R^{\textrm{min}}$ for  JTCN which is optimized using PCM-1 but evaluated with all PCMs. The goal of this analysis is to investigate whether used PCM for solving JTCN affects the conclusions and observed trends.

Fig.\,\ref{fig:EE_JT_opt_PCM1_eval_all_PCMsfig} suggests the following two observations. First, as expected, increasing the amount of power we account for in our PCM leads to a decrease in the energy efficiency. \textit{In other words, a simple PCM tends to overestimate the energy efficiency.} This overestimation is close to 
two orders 
of magnitude when the data rate requirement is low, e.g., $R^{\textrm{min}}=10$ Kbps. The overestimation decreases with higher rate requirements, however still is significant, e.g., around $2.39\times$ 
when $R^\textrm{{min}}=2$\,Mbps. This overestimation is particularly problematic, especially for battery-operated devices, as the device and consequently the network lifetime will be much shorter than the anticipated lifetime. 
Second, when optimizing using PCM-1 and evaluating with the same PCM~(line with $\diamond$ markers in Fig.\,\ref{fig:EE_JT_opt_PCM1_eval_all_PCMsfig}), we can observe energy efficiency as a decreasing function of $R^{\textrm{min}}$, which is not the case if evaluated with other PCMs. 
We expect this trend to emerge due to the lack of circuit power cost component in PCM-1 which leads the optimal operation point to be closer to the point where the minimum amount of power is allocated. When there is an offset cost like circuit power consumption, power allocation schemes prefer allocating higher transmission power to achieve higher throughput, which then can improve energy efficiency.  
Also, as will be shown in the next subsection, users who sustain a higher data rate than their $R^\textrm{{min}}$ are the users with the best channel quality in each BS.\footnote{Similar observations are reported in \cite{rezvani2022optimal}.}  
In a nutshell, the impact of a simple PCM is particularly pronounced on low $R^\textrm{{min}}$ regimes where the optimal mode of operation is more likely to be to serve these users with higher data rates, while for high $R^\textrm{{min}}$ values most of the available power is used to meet the rate requirement of the worst-channel-quality users. 
%


\begin{figure*}[t]\vspace{-10pt}
\subfloat[Outage ratio.]{\includegraphics[width=0.33\linewidth]{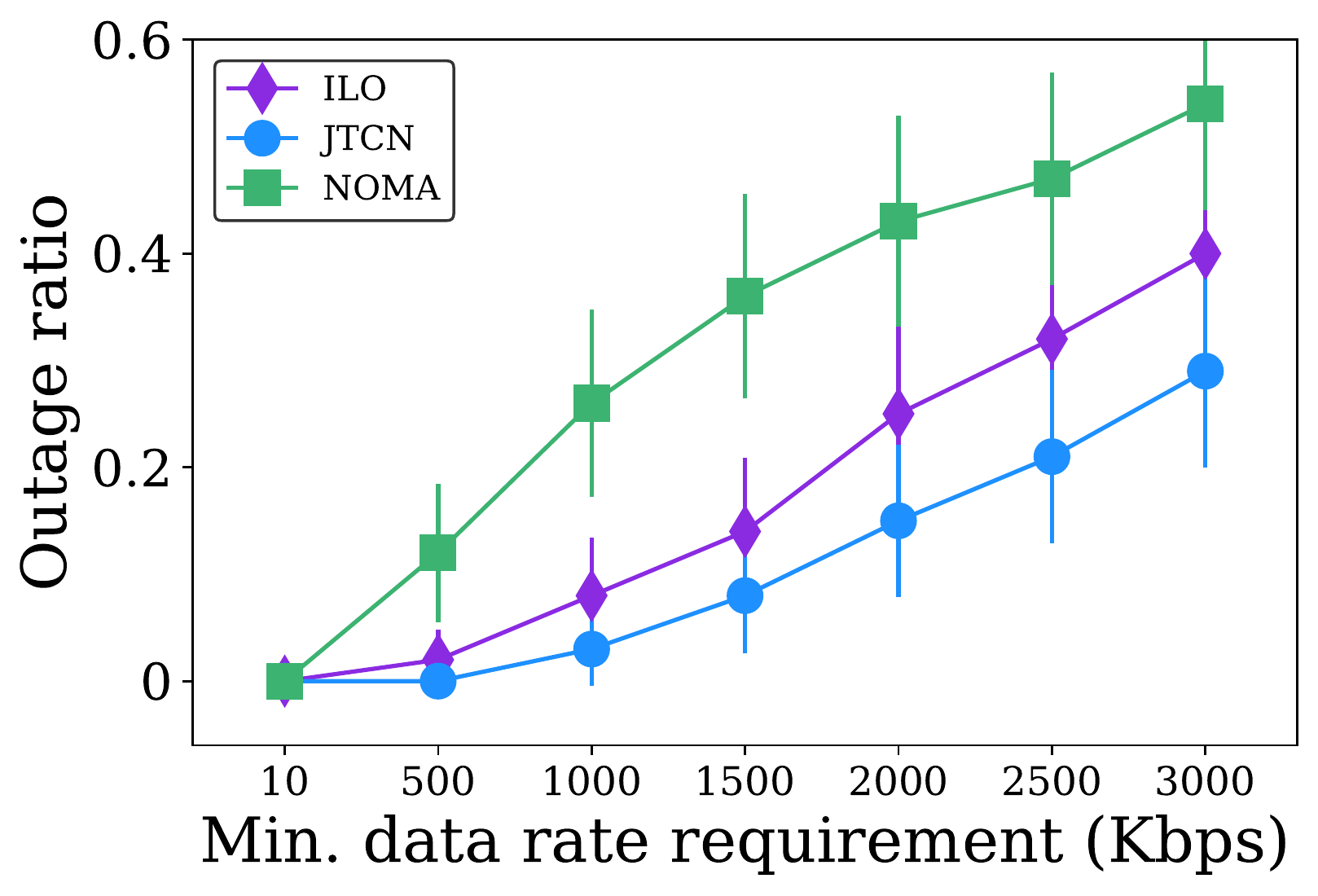}
\label{fig:outage_local_vs_global}}
\subfloat[Network energy efficiency.]{\includegraphics[width=0.33\linewidth]{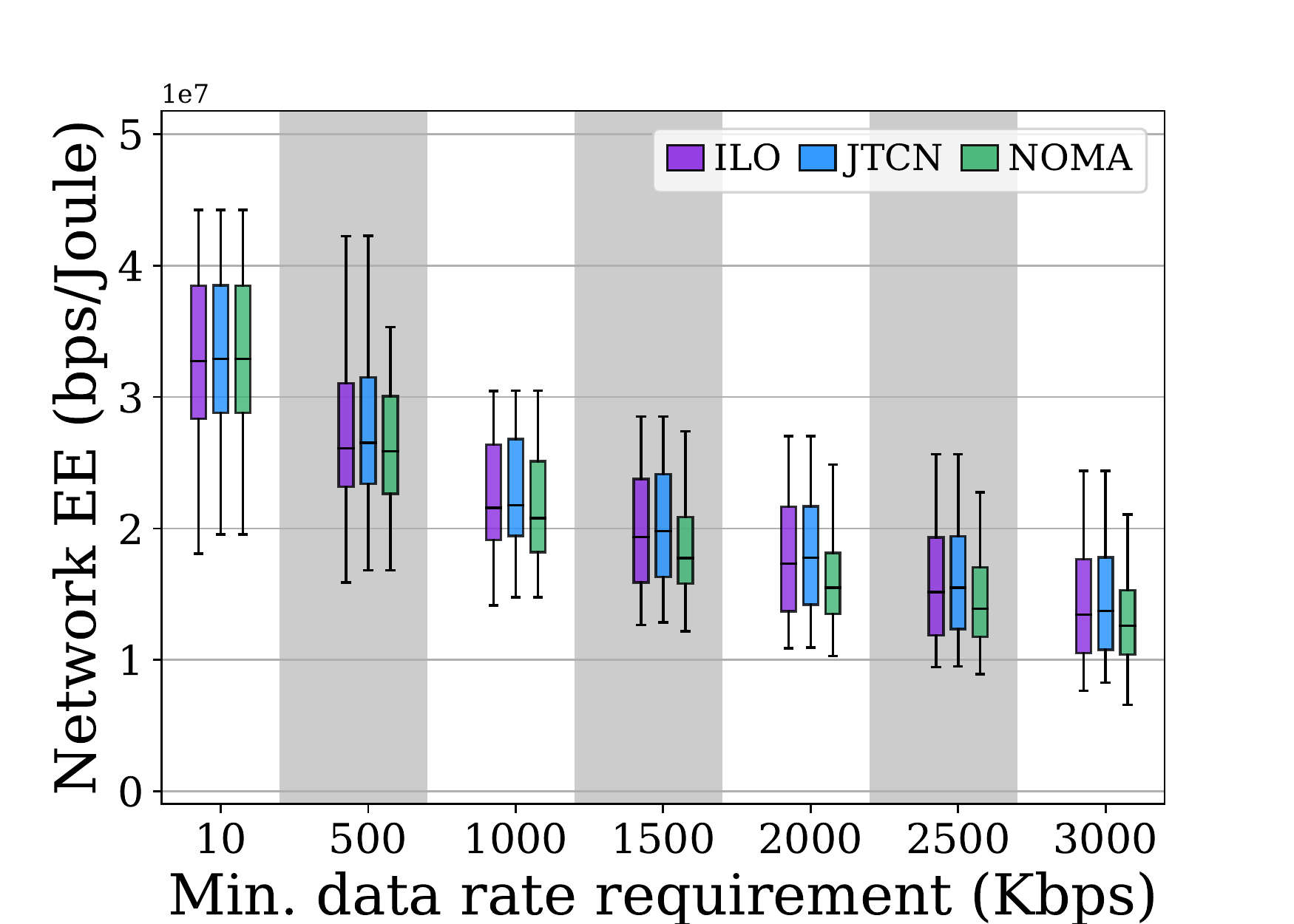}
\label{fig:ee_local_vs_global}} 
\subfloat[Throughput.]{\includegraphics[width=0.33\linewidth]{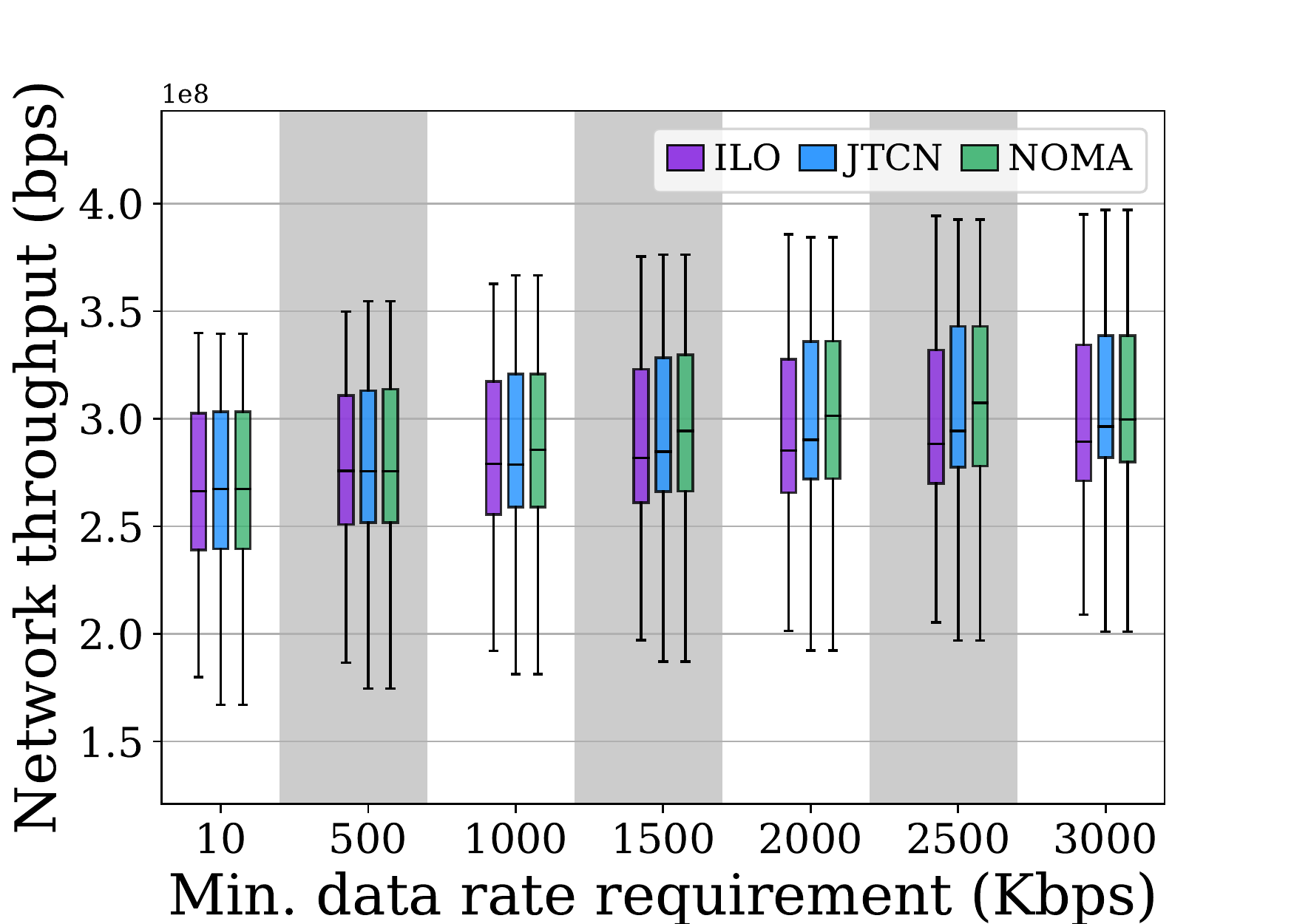} \label{fig:thru_local_vs_global}}%
  \caption{Comparison of JTCN and NOMA under increasing $R^{\textrm{min}}$ for global and ILO optimized and evaluated with PCM-$\kappa$. }%
  \label{fig:incrate_local_vs_global} \vspace{-10pt}
\end{figure*}

Fig.\,\ref{fig:TP_JT_opt_all_eval_PCMprop} shows the change in throughput achieved by JTCN under increasing $R^\textrm{{min}}$. In this case, JTCN scenario is optimized for each PCM while its energy efficiency is evaluated using only PCM-$\kappa$  as we believe that this offers a better approximation of the real-world power consumption compared to other PCMs. 
Moreover, the red line in Fig.\,\ref{fig:TP_JT_opt_all_eval_PCMprop} represents the minimum required throughput to guarantee that all users are served with at least their respective requested data rates. 
For low $R^\textrm{{min}}$ values, the optimization with PCM-1 allocates less power than necessary for optimizing the \textit{real} energy efficiency, since this PCM considers lower circuit power. For $R^\textrm{{min}}=10$ Kbps, optimizing for PCM proposed yields $3.1\times$ higher throughput compared to that of PCM-1. The corresponding  total power allocated by the two BSs is approximately $0.07$ W (0.18\% of the available power) when optimizing with PCM-1 while $1.49$ W (3.73\% of the available power) was used when optimizing with PCM proposed. 
The throughput difference is smaller for higher  data rate requirements, e.g., for $3$ Mbps the difference is only $7$\% for throughput and the average system allocated power is $16.13$ W (40.41\% of the available power) for PCM-1 and $16.65$ W (42.48\% of the available power) for PCM-$\kappa$.
In this case, 
each BS prefers to use more of its power budget to meet the rate requirements of its users, thereby resulting in approximately the same throughput for all schemes. Yet, not all power budget is exhausted by these schemes, as it would not yield the highest energy efficiency.
Note that all schemes maintain the network throughput  that is significantly higher than the throughput minimally needed for users' applications. 

When it comes to energy efficiency,  Fig.\,\ref{fig:EE_JT_opt_all_eval_PCMprop} suggests also that the impact of the PCM used for optimization becomes more visible in low $R^\textrm{{min}}$ regimes.  
For example, when $R^\textrm{{min}}=10$ Kbps, optimizing for PCM proposed yields $2.57\times$ higher energy efficiency compared to that of PCM-1, while this gain is only 3\% for $R^\textrm{{min}}=3$ Mbps. When comparing with PCM-2 we observe an even smaller difference, ranging from 1-6\%.

\begin{tcolorbox}[title=\textbf{Takeaway},colback=gray!20,colframe=white,arc=2mm,boxsep=3mm]
The accuracy of the PCM plays a key role in the achieved network energy efficiency, especially for low $R^\textrm{{min}}$ regimes. Given that NOMA is considered to be promising to serve low-rate IoT applications~\cite{yuan2021noma}, it is paramount to consider more realistic PCMs for such scenarios. In these regimes, the use of an oversimplified PCM in power allocation might lead to considerable loss in energy efficiency and throughput, while more accurate models like PCM-2 and our proposal can improve the energy efficiency.  Moreover, when it comes to the PCM used for performance evaluation, oversimplified PCMs might lead to optimistic energy efficiency values which diverge from the reality with several orders of magnitude, resulting in much shorter network and device lifetime than the anticipated optimistic lifetimes.

\end{tcolorbox}

\subsection{JTCN versus Conventional NOMA}
\label{subsec:JT_vs_Conv}

Now, let us compare the performance of JTCN against NOMA (global and ILO) considering PCM-$\kappa$.
Fig.\,\ref{fig:outage_local_vs_global} shows the outage ratio under increasing $R^\textrm{{min}}$. For low $R^{\textrm{min}}$, all schemes can find a feasible solution, whereas with increasing rate requirement, NOMA starts to result in many outages. This is due to the high interference experienced by the cell-edge user, whose signal quality is not strong enough for the required $R^{\textrm{min}}$. Meanwhile, JTCN suffers less from outage as it eliminates the interference on the cell-edge user by the two BSs jointly serving this user. However, we still observe a non-negligible fraction of cases not being feasible for NOMA even under JTCN.   
A big fraction of the outage cases experienced in NOMA can be eliminated by choosing the correct BS that is typically the BS closest to the cell-edge user. In our conventional NOMA scenario, we assume that the cell-edge user is connected to BS$_1$. This points to the importance of user association in the operation of NOMA.  

\begin{figure*}[ht]
\centering
\begin{minipage}[b]{.43\textwidth}
\centering
\includegraphics[width=1\textwidth]{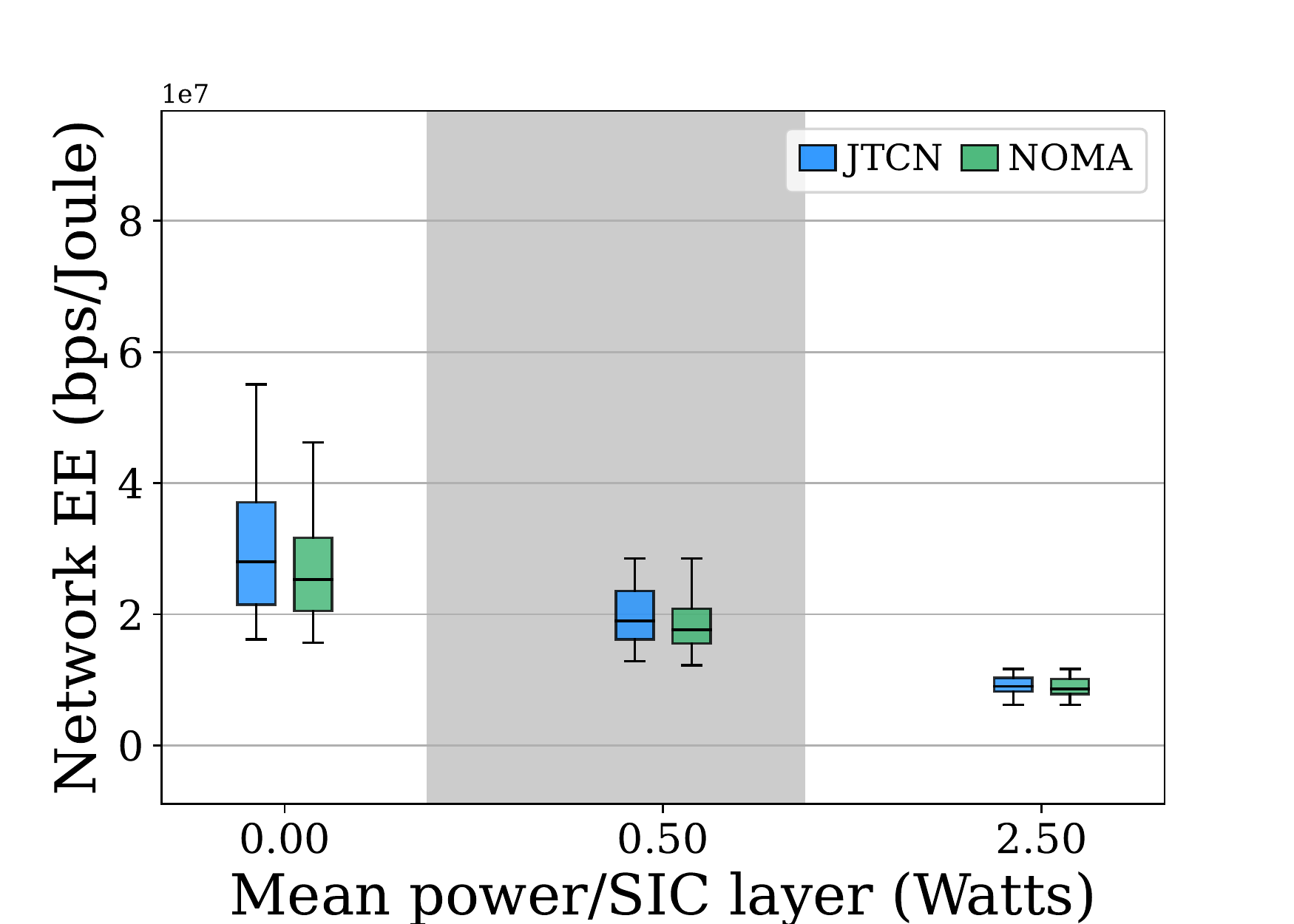}
\caption{EE of JTCN and NOMA optimized using PCM-$\kappa$ with $\kappa=\{0, 0.5, 2.5\}$ and  $R^{\textrm{min}}=1.5$ Mbps for all users. \label{fig:EE_var_kappa_R_1500}}
\end{minipage}
\hfill
\begin{minipage}[b]{.49\textwidth}
\centering
\includegraphics[width=1\textwidth]{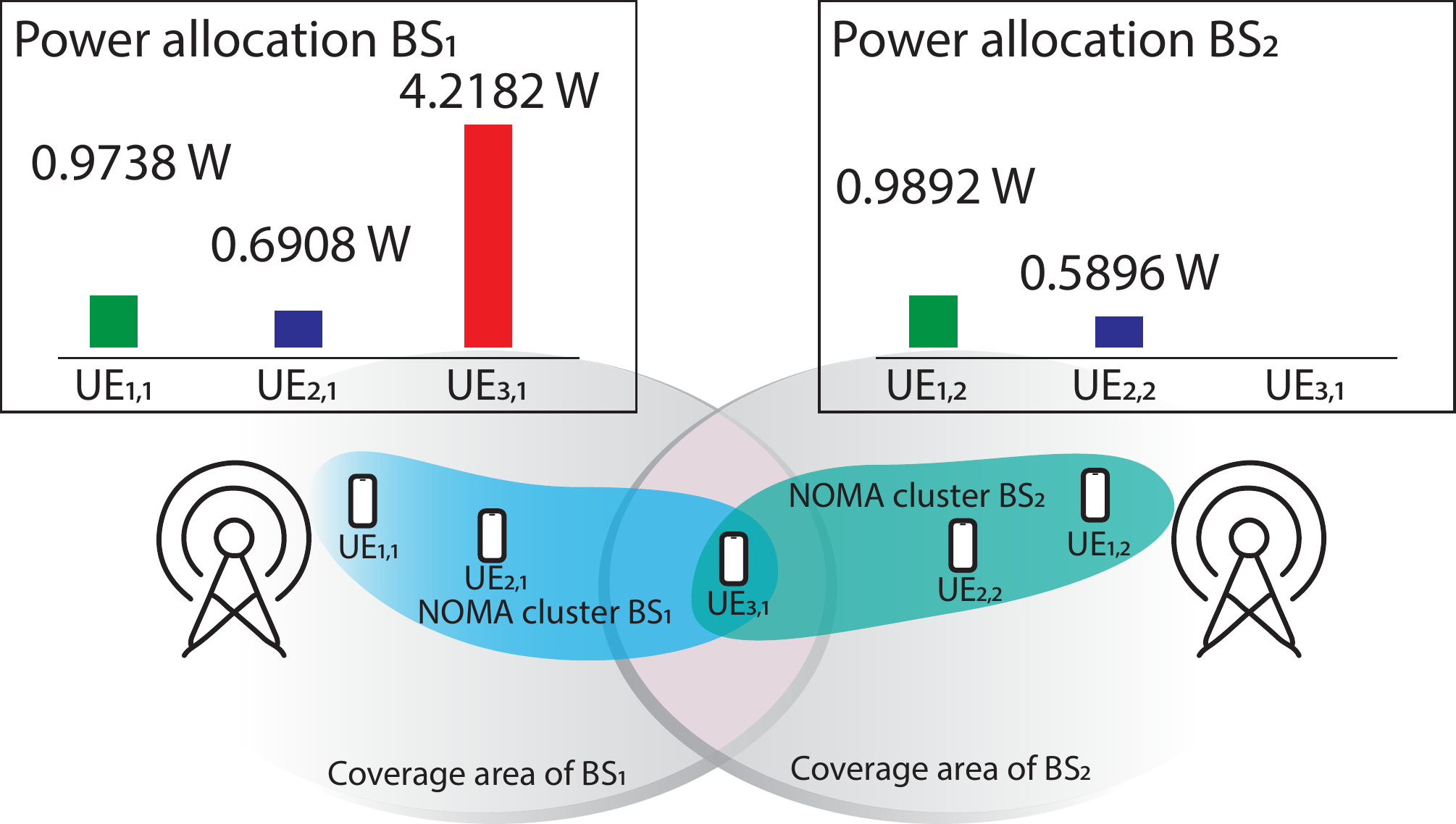}
\caption{Allocated power values when $R^{\textrm{min}}=$3 Mbps in JTCN. \label{fig:power_aloc_JT_example}}
\end{minipage}
\end{figure*}
We observe that for most of the cases the JTCN's optimal operation point is the same as \textit{dynamic cell selection (DCS)-CoMP}~\cite{shin2017non} in which the two BSs cooperate to serve the user, but only the BS with the best channel serves the user and the serving BS can be dynamically changed based on the channel quality. 
In other words, JTCN converges to conventional NOMA in most cases. 
But, JTCN leverages both BSs to jointly serve the cell-edge user 
only in case of high $R^\textrm{{min}}$ or during deep fading. \textit{Therefore, the key benefit of JTCN over conventional NOMA is the lower outage ratio. }
From Fig.\,\ref{fig:outage_local_vs_global}, we also observe that ILO yields a slightly higher outage ratio in comparison to JTCN while still providing benefits over NOMA. We attribute this trend to the fact that  the cell-edge user in ILO is served by the BS that offers the best channel condition. When it comes to the convergence of ILO, it is typically only 2-5 iterations.   
For the remaining analysis, to ensure a fair comparison, we consider only the cases where 
all schemes 
have a feasible solution, i.e., none experiences outage. This, however, leads to a biased sample.

Fig.\,\ref{fig:ee_local_vs_global} and \ref{fig:thru_local_vs_global} show, respectively, the average network energy efficiency and  throughput under increasing $R^{\textrm{min}}$ for the three scenarios considered in this work, i.e., JTCN, NOMA, and ILO. These figures suggest that although JTCN always performs equally or better than the other two schemes in terms of energy efficiency, the performance difference is surprisingly almost insignificant (ranging from 0.0007\% for $R^{\textrm{min}} = 10$ Kbps to 10\% for $R^{\textrm{min}} = 3$ Mbps compared to NOMA and up to 2\% difference when comparing with ILO). Some prior studies comparing JTCN and NOMA such as \cite{muhammed2021resource} reported significantly higher energy efficiency of JTCN, however, without considering SIC overhead. This contradictory observations might occur due to the simple PCMs used in the literature. Moreover, despite optimizing locally, ILO maintains almost the same level of performance as the global approach by simply selecting the correct BS for the cell-edge user. 

Now, let us explore whether our conclusion is valid only for a certain $\kappa$ value or we observe a similar trend for different $\kappa$ values.  
Fig.\,\ref{fig:EE_var_kappa_R_1500} shows the network energy efficiency for $\kappa=\{0, 0.5, 2.5\}$ Watts/SIC layer and considering $R^{\textrm{min}}=1.5$ Mbps for all users. Setting $\kappa = 0$ corresponds to the case where SIC overhead is negligible  
and only radiated power-dependent power expenditure is accounted for. 
Fig.\,\ref{fig:EE_var_kappa_R_1500} suggests the following two conclusions. First, the network EE decreases as a function of $\kappa$, which is expected since $\kappa$ is the power expenditure of each SIC layer. Second, in the case of higher $\kappa$ values, network energy efficiency achieved with JTCN and NOMA is similar. For example, when $\kappa=0$, JTCN has 10\% higher energy efficiency compared to NOMA. This gap decreases to 6\% when $\kappa=0.5$ and to 2\% when $\kappa=2.5$. For $\kappa=0.5$ the SIC power consumption represents on average 55\% for $R^{\textrm{min}}=10$ Kbps and 19\% for $R^{\textrm{min}}=3$ Mbps in JTCN. For $\kappa=2.5$ the SIC power consumption represents on average 76\% for $R^{\textrm{min}}=10$ Kbps and 52\% for $R^{\textrm{min}}=3$ Mbps in JTCN.



%



Now, let us focus on a specific instance and observe the resulting power allocation for each user under JTCN in Fig.\,\ref{fig:power_aloc_JT_example} for   $R^\textrm{{min}}=3$ Mbps. Although  BS$_1$ and BS$_2$ could serve UE$_3$ jointly, this user is being served only by the first BS: that is, JTCN converges to the conventional NOMA. In fact, for this snapshot of the network, the resulting allocated power for the JTCN or conventional NOMA is  the same. Similar results can be seen throughout all feasible samples. However, two situations lead to different allocations for the mentioned scenarios: (i) when the chosen BS to transmit to the cell-edge user is BS$_2$  in JTCN, and (ii) when the chosen BS is the same, but it is an infeasible problem for conventional NOMA. In this case, joint transmission is the only possible solution and the cell-edge user will be served by both BSs. 

\begin{tcolorbox}[title=\textbf{Takeaway},colback=gray!20,colframe=white,arc=2mm,boxsep=3mm]
Our numerical analysis suggests that there is not a significant energy efficiency benefit of JTCN over conventional NOMA. However, JTCN still provides benefits in terms of lower outage ratio, i.e., it can find feasible solutions to the power allocation problem when NOMA cannot find one due to the unmanaged interference from the neighboring cell(s). 
Surprisingly, JTCN converges to DPS-CoMP when there exists a feasible solution for conventional NOMA and serving by only a single BS is a better option than jointly. 
Finally, despite working locally, ILO can provide very similar performance to that of the global approach by selecting the best BS for the cell-edge user and converges only after a few iterations. 
\end{tcolorbox}



\begin{figure}[t]
    \centering
\includegraphics[width=0.45\textwidth]{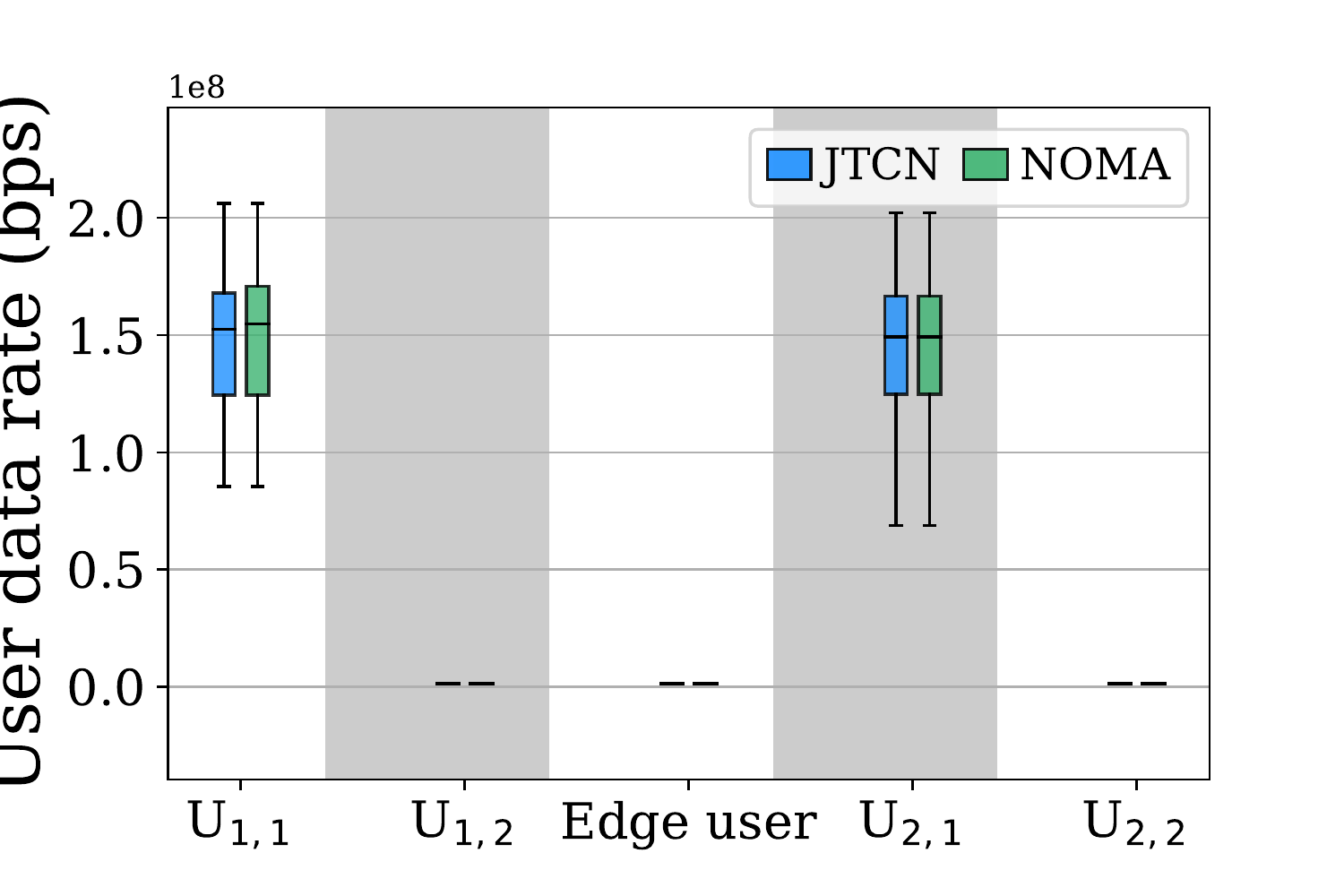}
    \caption{User data rate of JTCN and NOMA optimized using PCM-$\kappa$ for $R^{\textrm{min}}=1.5$ Mbps for all users.}
    \label{fig:TP_users_1500Kbps_PCMprop} 
\end{figure} 
\begin{figure*}
\subfloat[$\kappa = 0$]{\includegraphics[width= 0.33 \textwidth]{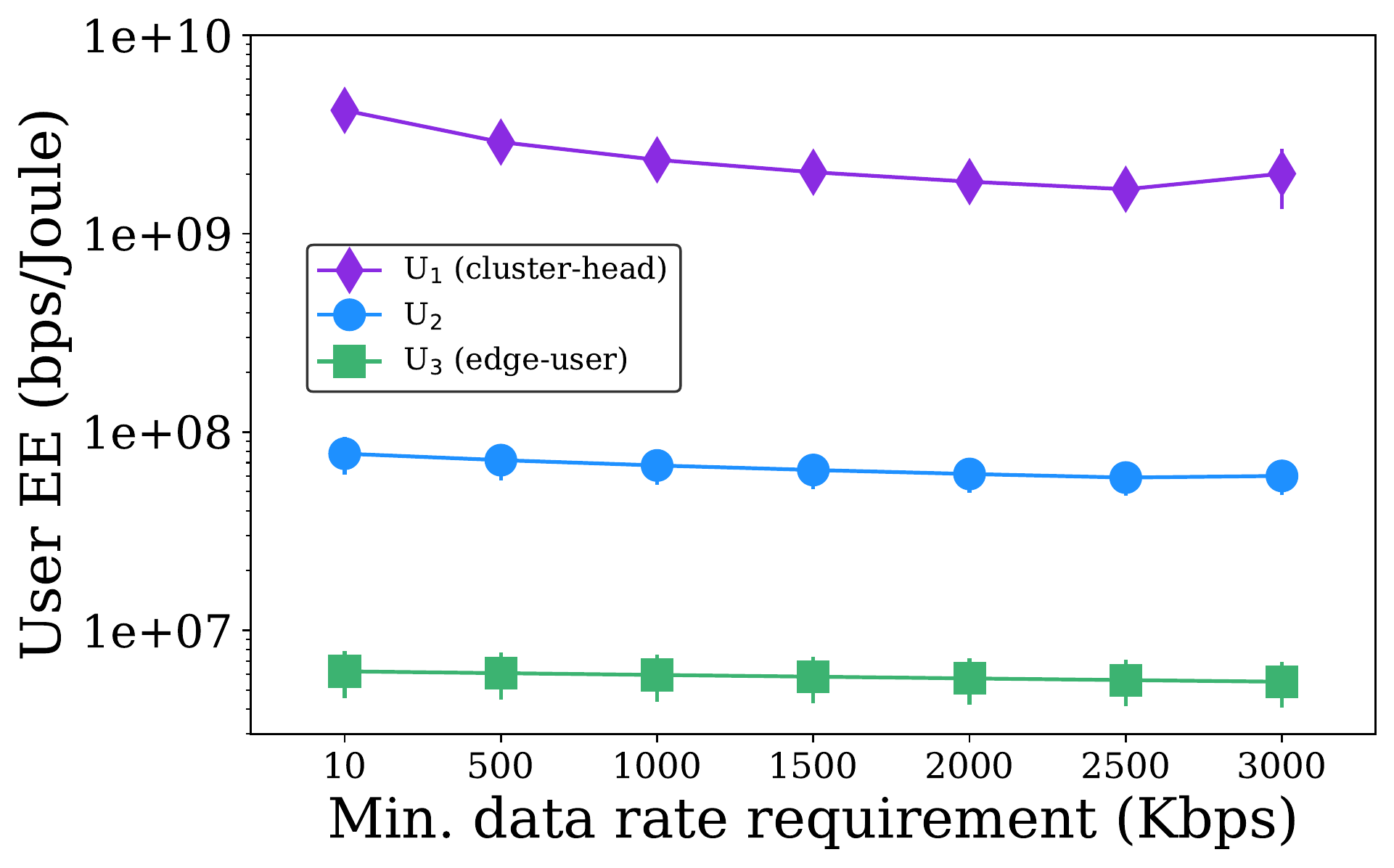} \label{fig:kappa0EEusers}}
\subfloat[$\kappa = 0.5$]{\includegraphics[width= 0.33 \textwidth]{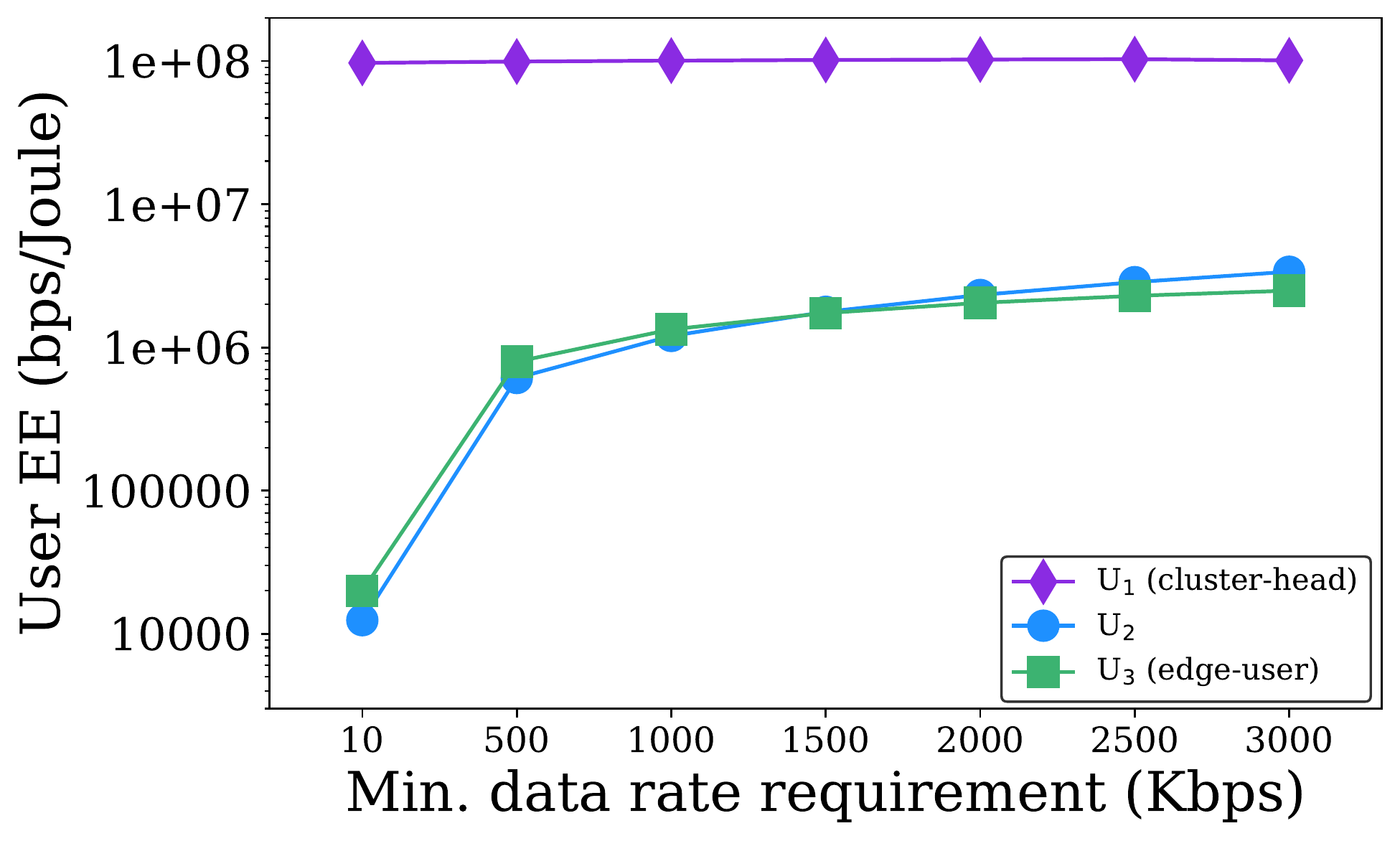}\label{fig:kappa05EEusers}}
\subfloat[$\kappa = 2.5$]{\includegraphics[width= 0.33 \textwidth]{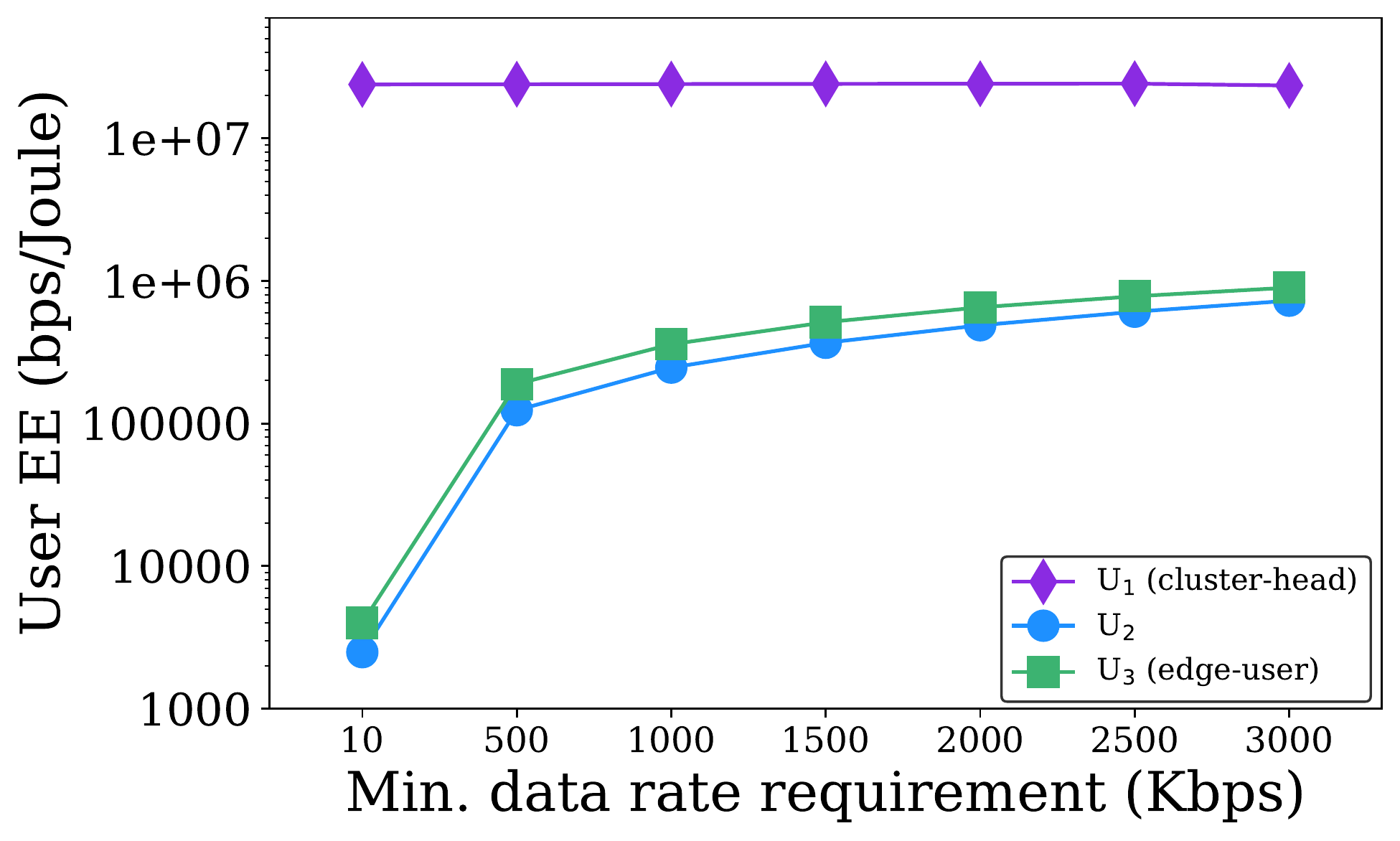}\label{fig:kappa25EEusers}}
\caption{Individual energy efficiency of the users in the coverage area of the first BS.\label{fig:EE_users}}
\end{figure*}

\subsection{Users Energy Efficiency}
\label{subsec:user_EE}

Since energy efficiency of the end devices is crucial for maintaining longer battery lifetime and NOMA incurs SIC overhead at the receivers, let us investigate how different users are affected in the considered schemes in terms of maintained throughput and energy efficiency.

Fig.\,\ref{fig:TP_users_1500Kbps_PCMprop} shows each user's throughput under JTCN and NOMA with $\kappa = 0.5$ and $R^{\textrm{min}}=1.5$ Mbps for all users. This figure suggests that there is no statistically significant difference between NOMA and JTCN also in terms of throughput, and both result in a similar throughput distribution among the users. Moreover, users with the best channel in each BS are allocated more power than the power value that would be sufficient to satisfy the minimum required rate. 
Apart from these users (also referred to as \textit{cluster heads}), remaining users maintain only the minimum required data rate, indicating an unfair throughput distribution. Hence, NOMA schemes optimizing for energy efficiency should also introduce a notion of fairness among users to mitigate this throughput difference between the cluster-head and other users. 

Now, let us investigate whether there exists also difference in the  energy efficiency of the users. Fig.\,\ref{fig:EE_users} shows the average energy efficiency of each user served by BS$_1$ under increasing $R^{\textrm{min}}$ considering JTCN for $\kappa=\{0, 0.5, 2.5\}$. Our observations hold also for NOMA, hence we omit those NOMA results in the figures for the sake of conciseness. 
Fig.\,\ref{fig:kappa0EEusers} shows an order of magnitude difference among the energy efficiency of the users. As previously observed, U$_1$ with the best channel maintains visibly higher energy efficiency with  two orders of magnitude difference to that of the cell-edge user. This difference is consistently observed under all minimum rate requirements and energy efficiency remains almost the same under all $R^{\textrm{min}}$ regimes.

In Fig.\,\ref{fig:kappa05EEusers} and Fig.\,\ref{fig:kappa25EEusers}, we observe a similar energy efficiency gap between the cluster-head user and the remaining users. In contrast to the case with $\kappa=0$, there is no significant performance difference between the cell-edge user and U$_2$. Moreover, energy efficiency of these two users incrase with increasing  $R^{\textrm{min}}$. We attribute this behaviour to the impact of $\kappa$; 
for $\kappa>0$, SIC power consumption might become more dominant factor in total power consumption. In this case, the optimal user energy efficiency for the non-cluster-head users is potentially in a region with higher data rate than their $R^{\textrm{min}}$. However, as aforementioned, since the optimization problem aims at maximizing the network energy efficiency rather than the users' energy efficiency, these users will only get their minimum required data rate, thereby maintaining a lower energy efficiency.  With the increase in $R^{\textrm{min}}$, these users get closer to their optimum point of operation, which explains the increase in energy efficiency for higher  $R^{\textrm{min}}$. Eventually, we observe a plateau for these users. Meanwhile, the cluster-head user's energy efficiency remains almost the same. Finally, the orders of magnitude difference in the achieved energy efficiency under different $\kappa$ regimes highlights the importance of accurately profiling the SIC energy consumption overhead. 



\begin{tcolorbox}[title=\textbf{Takeaway},colback=gray!20,colframe=white,arc=2mm,boxsep=3mm]
If fairness notion is not incorporated in power allocation problem, NOMA schemes might yield a significant throughput and energy efficiency difference among the users, in favour of the user with the best channel condition. This observation can also be considered while determining the NOMA clusters, e.g., users with high rate requirements can be put in the same NOMA cluster with nodes whose channels are weaker and rate requirements are lower. 
\end{tcolorbox}

\section{Discussion and Limitations}\label{sec:disc}


A limitation of our work is that similar to the previous works~\cite{choi2014non, tian2016performance}, due to the computational complexity of solving the formulated problems, we assessed the performance of our proposal in a small setting with a limited number of users. However, given that JT-CoMP coordination complexity increases with increasing number of BSs, we expect a small number of BSs jointly serving the edge-users. Similarly, due to the complexity of SIC, the literature also considers NOMA clusters of typically a few users, e.g., 2 users~\cite{muhammed2021resource} or 3 users~\cite{khan2019joint}. As future work, efficient heuristics can be designed to observe the impact of the PCMs and performance of JTCN in comparison to NOMA in larger settings.

Moreover, our PCM is a preliminary model which could incorporate other factors such as power amplifier's efficiency. Similarly, for JTCN, we have not accounted for the energy consumption due to coordination among the BSs. The models from prior research such as \cite{muhammed2021resource} can be incorporated into our problem for a more realistic accounting of the JTCN overhead. With such an energy overhead, JTCN might maintain a lower energy efficiency compared to conventional NOMA. Finally, as discussed in \cite{lopez2022survey}, realistic power consumption models that can maintain sufficient balance between accuracy and tractability are essential in the development of more energy-efficient solutions. Equipment manufacturers and network operators should have test frameworks to characterize such power consumption profiles and share these observed profiles transparently with the public for a more thorough understanding of the power consumption and consequently to design more energy-efficient schemes. Future work can consider more heterogeneous settings with users requiring different minimum rate requirements and more elaborate PCMs. 







\section{Conclusion}
\label{sec:conc}
This paper investigates energy efficiency of NOMA and JT-CoMP NOMA~(JTCN) considering the power consumption overhead due to SIC processing at the NOMA receivers. 
We formulated an energy-efficiency-maximizing power allocation problem for a downlink multicell NOMA network. Relying on the Dinkelbach's algorithm, we proposed two approaches: an iterative local algorithm running at a BS and a global approach running at a centralized entity. 
To explore the impact of the power consumption model~(PCM) used for estimating the expected energy efficiency, we also considered the most common PCMs in the literature. Our numerical investigation showed that, simplistic PCMs tend to overestimate the energy efficiency of the network and result in lower throughput and energy efficiency when users' rate requirements are lower. 
Comparing JTCN with conventional NOMA, we observe that difference in their energy efficiency and throughput performance is only marginal. But, JTCN can find feasible power allocation solutions in many more scenarios compared to conventional NOMA, reflected as lower outage ratio. Additionally, when there exists a feasible solution for conventional NOMA, simulation results show that the optimal solution for JTCN is not to use coordination, i.e., it converges to a conventional NOMA. Finally, local approach performs close to global approach and converges after a few iterations.

 

\appendices

\section{Lower bound on the data rate expressions}\label{ap:A}

\subsection{Lower bound for NOMA data rate expression}
We first introduce a lower bound on the system data rate using the following lower bound introduced in Lemma 4.1 of \cite{zappone2015energy}: 
\begin{equation}
    \log_2(1+\gamma) \geq a\log_2(\gamma) + c,
    \label{eq:lower_bound}
\end{equation}
\textrm{where $\gamma>0$ and $\gamma_0>0$, while $a$ and $c$ are defined as follows: } 
\begin{align} &a= \gamma_0/(1+\gamma_0) \textrm{ and } \label{eq:a}\\
& c = \log_2(1+\gamma_0)-\gamma_0\log_2(\gamma_0)/(1+\gamma_0).
\label{eq:c}
\end{align}
The inequality is tight with equality for $\gamma = \gamma_0$ and $\left. \dfrac{\partial\log_2(1+\gamma)}{\partial \gamma}\right|_{\gamma =\gamma_0} = \left. \dfrac{\partial(a\log_2(\gamma)+c)}{\partial \gamma}\right|_{\gamma =\gamma_0}$.

Therefore, we write the lower bound for the NOMA achievable data rate in (\ref{eq:NOMA_data_rate}) as follows:
\begin{align}
    R_{i,b}^{\textrm{NOMA}}
    & \geq a_{i,b}\omega B\log_2\left(\dfrac{ p_{i,b}\chgain_{i,b}}{ \sum\limits_{\substack{b'=1,\\ b'\neq b}}^{N_{BS}}\sum\limits_{j=1}^{J_{b'}} p_{j,b'}\chgain_{i,b,b'} + \sum\limits_{j=1}^{i-1} p_{j,b}\chgain_{i,b} +\omega} \right) + c_{i,b}\omega B \nonumber \\
    &= a_{i,b}\omega B\log_2\left(p_{i,b}\chgain_{i,b}\right) + c_{i,b}\omega B  - a_{i,b}\omega B\log_2\left( \sum\limits_{\substack{b'=1,\\ b'\neq b}}^{N_{BS}}\sum\limits_{j=1}^{J_{b'}} p_{j,b'}\chgain_{i,b,b'} + \sum\limits_{j=1}^{i-1} p_{j,b}\chgain_{i,b} +\omega \right) \nonumber\\
    & = \Tilde{R}_{i,b}^{\textrm{NOMA}}.
    \label{eq:NOMA_data_rate_lower_bound}
\end{align}


With (\ref{eq:NOMA_data_rate_lower_bound}), we can lower-bound the objective function of the problem (\ref{eq:conv_obj_func_EE}) as:
\begin{align}
\dfrac{\sum\limits_{b=1}^{N_{BS}}\sum\limits_{i=1}^{J_b} R_{i,b}^{\textrm{NOMA}}}{P(\mathbf{p})} \geq
\dfrac{\sum\limits_{b=1}^{N_{BS}}\sum\limits_{i=1}^{J_b} \Tilde{R}_{i,b}^{\textrm{NOMA}}}{P(\mathbf{p})},
\label{eq:conv_obj_func_EE_lower_bound}
\end{align}
where the equality holds when $\{a_{i,b}, c_{i,b}\}_{\forall i, \forall b}$ are computed using (\ref{eq:a}) and (\ref{eq:c}). However, the numerator of the lower-bound function (\ref{eq:conv_obj_func_EE_lower_bound}) is still a sum of a difference of convex functions.
To force the concavity of the numerator, we introduce the following variable substitution $p_{i,b} = 2^{q_{i,b}}$, which allows us to rewrite $\Tilde{R}_{i,b}^{\textrm{NOMA}}$ as follows:
\begin{align}
    a_{i,b}\omega B(\log_2\left(\chgain_{i,b}\right) + q_{i,b})+ c_{i,b}\omega B  - a_{i,b}\omega B\log_2\left( \sum\limits_{\substack{b'=1,\\ b'\neq b}}^{N_{BS}}\sum\limits_{j=1}^{J_{b'}} 2^{q_{j,b'}}\chgain_{i,b,b'} + \sum\limits_{j=1}^{i-1} 2^{q_{j,b}}\chgain_{i,b} +\omega \right).
    \label{eq:NOMA_data_rate_lower_bound_var_sub}
\end{align}

Consequently,  (\ref{eq:NOMA_data_rate_lower_bound_var_sub}) is concave in the new variable since the log-sum-exp function is convex \cite{boyd2004convex}.

\subsection{Lower bound for JT-CoMP NOMA data rate expression}
Let us start introduce the same lower bound in (\ref{eq:lower_bound}). Consequently,  (\ref{eq:NOMA_data_rate_CoMP_2}) is then lower bounded as follows: 
\begin{align}
   &R^{\textrm{JTCN}} = \omega B\log_2\left( 1 +  \dfrac{ \sum\limits_{b'=1}^{N_{BS}}p_{J_b,b'}\chgain_{J_b,b'} }{ \sum\limits_{b'=1}^{N_{BS}}\sum\limits_{j=1}^{J_{b'} - 1} p_{j,b'}\chgain_{J_b,b,b'} + \omega} \right)\nonumber\\
   &\geq a_{i,b}\omega B \log_2\left(\dfrac{ \sum\limits_{b'=1}^{N_{BS}}p_{J_b,b'}\chgain_{J_b,b'} }{ \sum\limits_{b'=1}^{N_{BS}}\sum\limits_{j=1}^{J_{b'} - 1} p_{j,b'}\chgain_{J_b,b,b'} + \omega} \right) + c_{i,b}\omega B \nonumber\\
   & = a_{i,b}\omega B \log_2\left(\sum\limits_{b'=1}^{N_{BS}}2^{q_{J_b,b'}}\chgain_{J_b,b'} \right) + c_{i,b}\omega B - a_{i,b}\omega B \log_2\left( \sum\limits_{b'=1}^{N_{BS}}\sum\limits_{j=1}^{J_{b'} - 1} 2^{q_{j,b'}}\chgain_{J_b,b,b'} + \omega \right). 
   \label{CoMP_data_rate_relaxation1}
\end{align}
The resulting lower bound (\ref{CoMP_data_rate_relaxation1}) is a difference of convex functions, which is not concave. To make it concave let us use Lemma 4.2 from \cite{zappone2015energy} and lower-bound (\ref{CoMP_data_rate_relaxation1}) as:
\begin{align}
    &(\ref{CoMP_data_rate_relaxation1}) \geq a_{i,b}\omega B \log_2\left(\prod\limits_{b'=1}^{N_{BS}}\left( \dfrac{2^{q_{J_b,b'}}\chgain_{J_b,b'}}{c^{(1)}_{b'}} \right)^{c^{(1)}_{b'}} \right) + c_{i,b}\omega B - a_{i,b}\omega B \log_2\left( \sum\limits_{b'=1}^{N_{BS}}\sum\limits_{j=1}^{J_{b'} - 1} 2^{q_{j,b'}}\chgain_{J_b,b,b'} + \omega \right) \nonumber\\
   &= a_{i,b}\omega B \sum\limits_{b'=1}^{N_{BS}}\log_2\left(\left( \dfrac{2^{q_{J_b,b'}}\chgain_{J_b,b'}}{c^{(1)}_{b'}} \right)^{c^{(1)}_{b'}} \right) + c_{i,b}\omega B  - a_{i,b}\omega B \log_2\left( \sum\limits_{b'=1}^{N_{BS}}\sum\limits_{j=1}^{J_{b'} - 1} 2^{q_{j,b'}}\chgain_{J_b,b,b'} + \omega \right) \nonumber\\
    &= a_{i,b}\omega B \sum\limits_{b'=1}^{N_{BS}}c^{(1)}_{b'}\left( q_{J_b,b'} + \log_2\left( \dfrac{\chgain_{J_b,b'}}{c^{(1)}_{b'}} \right) \right) + c_{i,b}\omega B  - a_{i,b}\omega B \log_2\left( \sum\limits_{b'=1}^{N_{BS}}\sum\limits_{j=1}^{J_{b'} - 1} 2^{q_{j,b'}}\chgain_{J_b,b,b'} + \omega \right) \nonumber\\&= \Tilde{R}^{\textrm{JTCN}}
   \label{CoMP_data_rate_relaxation2_step1}
\end{align}
where $\sum\limits_{b'=1}^{N_{BS}} c^{(1)}_{b'} = 1$ and the inequality becomes an equality if the following holds:
\begin{equation}
 c^{(1)}_{b'} = \dfrac{2^{q_{J_b,b'}}\chgain_{J_b,b'}}{\sum\limits_{b'=1}^{N_{BS}} 2^{q_{J_b,b'}}\chgain_{J_b,b'}},\ \forall b'. 
 \label{eq:c1_b}
\end{equation}
The resulting lower bound (\ref{CoMP_data_rate_relaxation2_step1}) is the sum of an affine and a concave function with respect to $\{q_{i,b}\}$, which is concave. The same lower bound can be used in the data rate constraint to obtain:

\begin{align} 
    &-\sum\limits_{b'=1}^{N_{BS}}q_{J_b,b'}c^{(1)}_{b'} -\sum\limits_{b'=1}^{N_{BS}}\log_2\left(\dfrac{\chgain_{J_b,b'}^{c^{(1)}_{b'}}}{c^{(1)}_{b'}}\right)  +\log_2\gamma^{min}_{J_b,b}\left(\sum\limits_{b'=1}^{N_{BS}}\sum\limits_{j=1}^{J_{b'} -1 } 2^{q_{j,b'}}\chgain_{J_b,b,b'} +\omega\right) \leq 0.
   \label{eq:JT_problem_rate_constraint_3} 
\end{align}

\balance
\bibliography{references} 
\bibliographystyle{ieeetr}

\end{document}